\documentclass[preprint,12pt]{elsarticle}

\def\BlackHat{\textsc{BlackHat}}
\def\Sherpa{{\rm SHERPA}}
\def\Comix{{\rm Comix}}
\def\Amegic{{\sc AMEGIC++}}
\def\Root{\textsc{Root}}
\def\ntuple{$n$-tuple}
\def\nTuple{$n$-Tuple}

\def\ntuples{$n$-tuples}
\def\varref#1{{\tt #1}}

\usepackage{placeins}

\usepackage{amssymb}
\usepackage{color}
\usepackage{listings}
\usepackage{hyperref}
\usepackage{amsmath}
\newcounter{bla}

\def\sect#1{section~{\ref{#1}}}
\def\eqn#1{eq.~(\ref{#1})}

\def\Tab#1{Table~{\ref{#1}}}

\def\pT{p_{\rm T}}
\def\kT{k_{\rm T}}
\def\ET{E_{\rm T}}
\def\CC{{C\nolinebreak[4]\hspace{-.05em}\raisebox{.4ex}{\scriptsize{++}}}}
\def\SISCone{{\sc SISCone}}

\def\HTpartonicp{{\hat H}_{\rm T}'}
\def\HT{H_{\rm T}}

\def\Zj{$Z\,\!+\,1$}
\def\Zjj{$Z\,\!+\,2$}
\def\Zjjj{$Z\,\!+\,3$}

\def\Zgamjn{$Z,\gamma^*\,\!+\,n$}

\def\Zgamjx{$Z,\gamma^*\,\!+\,1,2$}
\def\Zgamjjjx{$Z,\gamma^*\,\!+\,3,4$}
\def\Zgamjjjxx{$Z,\gamma^*\,\!+\,1,2,3$}

\def\Wjn{$W\,\!+\,n$}
\def\Wj{$W\,\!+\,1$}
\def\Wjj{$W\,\!+\,2$}
\def\Wjjj{$W\,\!+\,3$}
\def\Wjjjj{$W\,\!+\,4$}

\def\Wjjjx{$W\,\!+\,1,2,3$}
\def\Wjjjjx{$W\,\!+\,1,2,3,4$}

\def\Wmjn{$W^-\,\!+\,n$}

\def\Wpjn{$W^+\,\!+\,n$}

\def\nn{\nonumber}
\def\jet{{\rm jet}}
\def\ETsl{{\s E}_{\rm T}}

\newbox\charbox
\newbox\slabox
\def\s#1{{      % Feynman slash
        \setbox\charbox=\hbox{$#1$}
        \setbox\slabox=\hbox{$/$}
        \dimen\charbox=\ht\slabox
        \advance\dimen\charbox by -\dp\slabox
        \advance\dimen\charbox by -\ht\charbox
        \advance\dimen\charbox by \dp\charbox
        \divide\dimen\charbox by 2
        \raise-\dimen\charbox\hbox to \wd\charbox{\hss/\hss}
        \llap{$#1$}
}}

\journal{Computer Physics Communications}
\newif\ifdraftnote
\draftnotetrue

\begin{document}

\begin{frontmatter}



\title{
{\normalsize \noindent
UCLA/13/TEP/108 \hfill
SLAC--PUB--15739\hfill
SB/F/422-13}\\
{\normalsize \noindent
IPhT--T13/228\hfill
\null\hskip 15mm IPPP-13-86\hfill
CERN--PH--TH/2013-243
}\\
$\null$ \\
Ntuples for NLO Events at Hadron Colliders}

%\author[a]{Z.~Bern\corref{zvi}}
\author{Z.~Bern\fnref{a}}
%\author[b]{L.~J.~Dixon\corref{lance}}
\author[b]{L.~J.~Dixon}
%\author[c]{F.~Febres Cordero\corref{fernando}}
\author[c]{F.~Febres Cordero}
%\author[b]{S.~H\"{o}che\corref{stefan}}
\author[b]{S.~H\"{o}che}
%\author[a]{H.~Ita\corref{harald}}
\author[d]{H.~Ita}
%\author[d]{D.~A.~Kosower\corref{david}}
\author[e]{D.~A.~Kosower}
\author[f,g]{D.~Ma\^{\i}tre\corref{daniel}}
%\author[e,f]{D.~Ma\^{\i}tre}

%\fntext[a]{??D}

%\cortext[zvi] {Corresponding author.\\\textit{E-mail address:} bern@physics.ucla.edu}
%\cortext[lance] {Corresponding author.\\\textit{E-mail address:} lance@slac.stanford.edu}
%\cortext[fernando] {Corresponding author.\\\textit{E-mail address:} ffebres@usb.ve}
%\cortext[stefan] {Corresponding author.\\\textit{E-mail address:} shoeche@slac.stanford.edu}
%\cortext[harald] {Corresponding author.\\\textit{E-mail address:} harald.ita@physik.uni-freiburg.de }
%\cortext[david] {Corresponding author.\\\textit{E-mail address:} David.Kosower@cea.fr}
\cortext[daniel] {Corresponding author. \textit{E-mail address:} daniel.maitre@durham.ac.uk}
%\cortext[kemal] {Corresponding author.\\\textit{E-mail address:} ozeren@physics.ucla.edu}

\address[a]{Department of Physics and Astronomy, UCLA, Los Angeles, CA 90095-1547, USA}
\address[b]{SLAC National Accelerator Laboratory, Stanford University, Stanford, CA 94309, USA}
\address[c]{Departamento de F\'{\i}sica, Universidad Sim\'on Bol\'{\i}var, Caracas 1080A, Venezuela}
\address[d]{Physikalisches Institut, Albert-Ludwigs-Universit\"at Freiburg, D-79104 Freiburg, Germany}
\address[e]{Institut de Physique Th\'eorique, CEA--Saclay, F--91191 Gif-sur-Yvette cedex, France}

\address[f]{Theory Division, Physics Department, CERN, CH--1211 Geneva 23, Switzerland}
\address[g]{Department of Physics, University of Durham, Durham DH1 3LE, UK}

\begin{abstract}
We present an event-file format for the dissemination of 
next-to-leading-order (NLO) predictions for QCD processes at hadron colliders.
The files contain all information required to compute generic
jet-based infrared-safe observables at fixed order (without showering
or hadronization), and to recompute observables
with different factorization and renormalization scales.  The files also 
make it possible to evaluate cross sections and distributions with
different parton distribution functions.   This in turn makes it possible
to estimate uncertainties in NLO predictions of a wide variety of
observables without recomputing the short-distance matrix elements.  The event
files allow a user to choose among a wide range of commonly-used jet algorithms and
jet-size parameters.

We provide event files for a $W$ or $Z$ boson accompanied by up to four jets, 
and for pure-jet events with up to four jets.  The files are for the Large 
Hadron Collider with a center of mass energy of 7 or 8 TeV.

A \CC{} library along with a Python interface for handling these files
are also provided and described in this article.  The library allows a
user to read the event files and recompute observables transparently for different
pdf sets and factorization and renormalization scales.
 \end{abstract}


\end{frontmatter}

% \linenumbers

% Computer program descriptions should contain the following
% PROGRAM SUMMARY.

{\bf PROGRAM SUMMARY}
  %Delete as appropriate.

\begin{small}
\noindent
{\em Manuscript Title: }                                       \\
{\em Authors:}                                                \\
{\em Program Title:}                                          \\
{\em Journal Reference:}                                      \\
  %Leave blank, supplied by Elsevier.
{\em Catalogue identifier:}                                   \\
  %Leave blank, supplied by Elsevier.
{\em Licensing provisions:}                                   \\
  %enter "none" if CPC non-profit use license is sufficient.
{\em Programming language:} \CC{} with Python interface       \\
{\em Computer:}                                               \\
  %Computer(s) for which program has been designed.
{\em Operating system: Linux, MacOS}                           \\
  %Operating system(s) for which program has been designed.
{\em RAM: varying} bytes                                        \\
  %RAM in bytes required to execute program with typical data.
%{\em Number of processors used: 1}                              \\
  %If more than one processor.
%{\em Supplementary material:}                                 \\
  % Fill in if necessary, otherwise leave out.
  % Please give some freely chosen keywords that we can use in a
  % cumulative keyword index.
{\em Classification:}                                         \\
  %Classify using CPC Program Library Subject Index, see (
  % http://cpc.cs.qub.ac.uk/subjectIndex/SUBJECT_index.html)
  %e.g. 4.4 Feynman diagrams, 5 Computer Algebra.
{\em External routines/libraries:} \Root, {\sc lhapdf\/}                      \\
  % Fill in if necessary, otherwise leave out.
%{\em Subprograms used:}                                       \\
  %Fill in if necessary, otherwise leave out.
%{\em Catalogue identifier of previous version:}*              \\
  %Only required for a New Version summary, otherwise leave out.
%{\em Journal reference of previous version:}*                  \\
  %Only required for a New Version summary, otherwise leave out.
%{\em Does the new version supersede the previous version?:}*   \\
  %Only required for a New Version summary, otherwise leave out.
{\em Nature of problem:} NLO QCD predictions for vector boson + jets
 and jet processes for generic observables.
%, factorization and renormalization
%scales and parton distributions
\\
  %Describe the nature of the problem here.
   \\
{\em Solution method:} Event files\\
  %Describe the method solution here.
   \\
%{\em Reasons for the new version:}*\\
  %Only required for a New Version summary, otherwise leave out.
   \\
%{\em Summary of revisions:}*\\
  %Only required for a New Version summary, otherwise leave out.
   \\
%{\em Restrictions:}\\
  %Describe any restrictions on the complexity of the problem here.
   \\
%{\em Unusual features:}\\
  %Describe any unusual features of the program/problem here.
   \\
%{\em Additional comments:}\\
  %Provide any additional comments here.
   \\
%{\em Running time:}\\
  %Give an indication of the typical running time here.
   \\

%\begin{thebibliography}{0}
%\bibitem{1}Reference 1         % This list should only contain those items referenced in the                 
%\bibitem{2}Reference 2         % Program Summary section.   
%\bibitem{3}Reference 3         % Type references in text as [1], [2], etc.
                               % This list is different from the bibliography at the end of 
                               % the Long Write-Up.
%\end{thebibliography}
%* Items marked with an asterisk are only required for new versions
%of programs previously published in the CPC Program Library.\\
\end{small}

\section{Introduction}
\label{Introduction}

With last year's discovery of a Higgs-like
boson~\cite{AtlasHiggs,CMSHiggs} at the Large Hadron Collider (LHC),
the ATLAS and CMS experiments at CERN have rounded out our knowledge
of the particle content of the Standard Model.  Direct searches for
new physics beyond it, and indirect searches via precision
measurements of the properties of the Higgs-like boson and of the top
quark, remain as challenges for ongoing research.  Both of these
avenues beyond the Standard Model require extensive calculations in
QCD to next-to-leading order (NLO), to high jet multiplicity.  NLO
is the first order in perturbation theory to provide a quantitatively
reliable estimate of backgrounds due to Standard-Model processes~\cite{LesHouches08, LesHouches09}.

NLO calculations require the computation of both virtual and
real-emission corrections to processes with high jet multiplicity.
The former require the evaluation of one-loop corrections to the basic
tree-level process, and the latter require the integration of matrix
elements with an additional emitted parton over the phase space for the
additional emission.  Both parts of the correction are computationally
intensive.  In order to obtain sufficiently small statistical
uncertainties, one must typically evaluate the matrix elements at
millions of different kinematic points when performing phase-space
integrals by Monte Carlo sampling techniques.

The end results of the calculations are total cross sections or
differential distributions for various observables.  In developing an
analysis\footnote{We use this term in the sense used by
  experimenters when analyzing data.}, one may wish to evaluate
these quantities with different choices of experimental cuts.
Different cuts may also be required for comparison with different
experimental analyses.  The estimation of uncertainties due to
variation of the unphysical renormalization and factorization scales,
and due to our imperfect knowledge of the proton's parton distribution
functions (pdfs), also require that we we recompute the same
differential distributions many times.  The latter alone can require
dozens or grosses of evaluations.  A naive rerunning of the entire
calculation would force the computationally expensive short-distance
matrix elements to be recomputed a comparable number of times.

The actual evaluation of multiple differential distributions,
given a list of kinematic points and matrix-element weights,
is however computationally
relatively cheap.  It is therefore very desirable to amortize the
expensive task of matrix-element computation over many evaluations of
differential distributions.  We can do this by storing each of the
phase-space points, along with the matrix elements and other
information, in data files.  In order to compress these sizeable
volumes of data, we store them as \Root{} \ntuple{} files, or simply
`\ntuple{} files' for short.  We produce \ntuple{} files using
the \BlackHat{} one-loop library~\cite{BH}
 in conjuction with \Sherpa{}~\cite{sherpa1,sherpa2,AmegicI,AmegicII,Comix}.
In this paper we describe in more detail
an implementation of this strategy for a class of interesting
multi-jet final states at the Large Hadron Collider (LHC).  We
document \ntuple{} files for processes producing a single electroweak
vector boson ($W^\pm$ or $Z$) in association with 1, 2, 3, or 4
jets, and files for pure-QCD processes producing 2, 3, or 4 jets.  In
addition, we supply a \CC{} library that provides an interface to the
files.  The \ntuple{} files also allow other researchers, and
especially experimenters, to perform their own NLO analyses, using
refined cuts, without the computational expense and management
complexity of running high-multiplicity codes.  The computations
recorded in the \ntuple{} files do not include showering or
hadronization, and the events are {\it not\/} a suitable starting
point for matched showers with existing implementations.  In contrast
to files containing showered events, the \ntuple{} files documented
here have a relatively small number of parton momenta recorded per
event, which keeps file sizes manageable.
A reweighting technique for estimating scale and pdf uncertainties has
also been applied to particle-level event samples that have been
generated by merging NLO calculations with a parton
shower~\cite{aMC@NLO-uncertainties}.

The paper is organized as follows. In
section~\ref{NLOCalculationsSection}, we discuss the organization of
an NLO calculation.  In section~\ref{nTupleFileSection}, we present
the content of event files.  In section~\ref{NTupleUseSection} we
explain how to compute differential cross sections using the event
files.  In section~\ref{ChangingScalesSection}, we discuss additional
details needed to change scales or parton distribution functions from
those in the original calculation.  In section~\ref{LibrarySection},
we present a software library for reading event files.  We summarize
in section~\ref{Conclusions}.

\section{NLO Calculations}
\label{NLOCalculationsSection}

We wish to store and reuse the matrix-element information computed during
an NLO calculation.  In order to explain what information we need to record,
we first review the components of an NLO calculation.  We further refine 
the information needed in \sect{ChangingScalesSection}.

\def\Obs{{\cal O}}
\def\obs{v}
\def\LIPS{{\rm LIPS}}
\def\muf{\mu_{\rm F}}
\def\mur{\mu_{\rm R}}
\def\mufz{\mu_{{\rm F},0}}
\def\murz{\mu_{{\rm R},0}}
\def\murf{\mu_{\rm R,F}}
\def\LO{{\rlap{\scriptsize\rm LO}{}}}
\def\NLO{{\rlap{\scriptsize\rm NLO}{}}}
\def\Born{{\rm Born}}
\def\virt{{\rm virt}}
\def\finvirt{{\rm fin.{}\;virt}}
\def\real{{\rm real}}
\def\sigmah{\hat\sigma}
Let us consider a process that produces $n$ final-state objects (jets or
electroweak bosons).
At leading order (LO) in QCD, the prediction for associated observables would
be given in terms of a tree-level matrix element for a process scattering two colored
partons into $n$ final-state partons or electroweak bosons.  Schematically, at this
order we can write the differential cross-section in an observable $\Obs$ as follows,
\begin{equation}
\begin{array}{rl}
\displaystyle\frac{d\sigma^\LO}{d\obs} &= 
\displaystyle\int d\sigmah_n\,\delta_\obs\\
&= \displaystyle\int dx_1 dx_2 \int d\LIPS_n\; f_1(x_1) f_2(x_2) \,
   \sigmah_n \,\delta\bigl(\obs-\Obs(\{k\}_n)\bigr)\,,
\end{array}
\label{eq:LO}
\end{equation}
where $f_{1}(x_{1})$ and $f_{2}(x_{2})$ are the pdfs for the partons inside the
two incoming protons, $d\LIPS_n$ is the $n$-particle Lorentz-invariant
phase-space measure for momenta $\{k\}_n$, and $\sigmah_n$ is the
short-distance $2\rightarrow n$ squared matrix element.
We sum over all allowed parton species implicitly.  The pdfs
$f_{1,2}$ also depend on a factorization scale $\muf$, and the
parton-level squared matrix element $\sigmah_n$ depends on a
renormalization scale $\mur$.  These two scales will in general depend
on the phase-space point $\{k\}_n$.  However, at LO $\sigmah_n$ depends on
the renormalization scale only through its (homogeneous) dependence on
the strong coupling $\alpha_s(\mur)$.

In order for \eqn{eq:LO} to produce a meaningful answer, the phase
space must be cut off in regions where the squared matrix elements are
singular.  We are primarily interested in jet cross sections, and so
we shall use a jet algorithm to impose such cuts.  We leave them
implicit in \eqn{eq:LO}.

The integral will typically be performed by Monte-Carlo integration,
using a random sample of phase-space points.  More sophisticated
approaches will adapt the sample to the squared matrix element.  The
evaluation will generate a list of phase-space points and momentum
fractions, along with the associated integration weight.  At each
phase-space point, we evaluate only a single subprocess, fixing the
incoming and outgoing parton types and flavors.  In order to recompute
any observable in an LO calculation, or compute a new one, it then
suffices to save the generated phase-space configurations, along with each
configuration's integration weight.  Each configuration can be saved
by saving the four-momenta of each final-state parton it contains.  In
order to estimate the scale sensitivity of observables, we must
recompute them with different choices of renormalization scale.  For
this purpose, it suffices to save the choice of $\mur$ for each
configuration, as the weight is homogeneous in $\alpha_s$ with a fixed
power.  In order to estimate the uncertainty in an observable due to
imprecise knowledge of the pdfs, we must recompute them with different
pdfs drawn from an error ensemble.  For this purpose, we must also
save the initial-state momentum fractions and choice of $\muf$ for
each configuration.  It is also convenient to save separately the
matrix-element weight multiplied by the phase-space volume element;
this is the quantity that would yield the integration weight upon
multiplication by the pdfs.

The NLO prediction for the same process is more intricate.  We can
decompose the computation of an observable into a Born-level
contribution, along with virtual and real-emission corrections,
\begin{equation}
  \frac{d\sigma^\NLO}{d\obs}\;\; = \int\left(d\sigmah_n^\Born+d\sigmah_n^\virt\right)\delta_\obs
  +\int d\sigmah_{n+1}^\real\,\delta_\obs\,,
\end{equation}
where the subscripts indicate the number of particles in the
final-state phase space.  The virtual and the real-emission
corrections are separately infrared divergent. The virtual
corrections, given by the interference of one-loop and tree matrix
elements, have explicit divergences arising from the integration over
the loop momenta.  These are usually regulated using dimensional
regularization ($D=4-2\epsilon$), and appear as double and single
poles in $\epsilon$.  The infrared divergences in the real-emission
contributions arise from regions of $(n+1)$-body phase space in which
two massless partons become collinear or a gluon becomes soft.  We can
think of the $(n+1)$-body phase space factorizing into a phase space for $n$
final-state objects and an unresolved phase space over soft or
collinear partons.  Were we to integrate these contributions over a
dimensionally-regulated unresolved phase space, the resulting poles in
$\epsilon$ would cancel against those in the virtual contributions for
infrared-safe observables, allowing us to obtain a finite result in
the $\epsilon\rightarrow0$ limit.

However, the integration over the dimensionally-regulated
real-emission phase space is intractable analytically in the presence
of kinematical cuts.  Accordingly, we must integrate numerically; but
it is hard to extract singularities, and more importantly, the
underlying finite term, by numerical integration in $D$ dimensions.
Instead, we seek to separate the real-emission contributions into two
parts: a simple part, containing all singularities, to be integrated
analytically in $D$ dimensions; and a remainder, whose numerical
integral is finite over the complete phase space.  The universality of
the singular limits of the matrix elements makes this possible.  The
most common approach, which we use as well, is to add and subtract an
approximation to the real-emission squared matrix element that
captures all of its divergent collinear and soft limits, and yet is
simple enough to be integrated analytically over the unresolved
phase-space.  This approach started with applications to specific
processes~\cite{EllisRossTerrano, ManganoNasonRidolfi, KunsztSoper,
  FKS} and was later generalized by Catani and
Seymour~\cite{CataniSeymourShort, CataniSeymourLong} to a
process-independent method.  We use an implementation of the
Catani--Seymour dipole method within the \Amegic{}~\cite{AmegicI,AmegicII}
and \Comix{}~\cite{Comix} tree-level
matrix-element package (both parts of
\Sherpa~\cite{sherpa1,sherpa2}, with \Comix{} used for
the highest-multiplicity processes).  Other subtraction methods have also
been applied, especially in NNLO
calculations~\cite{FKS,TwoLoopAntenna}, and related methods are under
development~\cite{NagySoperSubtraction}.  Some of these have
also been automated~\cite{MadFKS}.
With a subtraction term, we
obtain the following formula for a differential cross section computed
to NLO,
\begin{equation}
\frac{d\sigma^\NLO}{d\obs}\;\; = \int d\sigmah_n^\Born\,\delta_\obs
+ \int \left(d\sigmah_n^\virt+d\sigmah_n^{\rm int}\right)\delta_\obs
 +\int\left(d\sigmah_{n+1}^{\rm real}-d\sigmah_{n+1}^{\rm sub}\right) \delta_\obs\,,
\label{eq:NLOfull0}
\end{equation}
where $\sigmah_n^{\rm int}$ is the integral of $\sigmah_{n+1}^{\rm
  sub}$ over the unresolved phase space.  Each integral in
\eqn{eq:NLOfull0} is now finite as $\epsilon\rightarrow0$.  (The
observable must be infrared and collinear safe, that is $\Obs_{n+1}$
must approach $\Obs_n$ in every singular limit, for this to be true.)
As the divergent terms in $\sigmah_n^\virt$ cancel those in
$\sigmah_n^{\rm int}$, we can drop them all, retaining only the finite
contributions,
\begin{equation}
\frac{d\sigma^\NLO}{d\obs}\;\; = \int d\sigmah_n^\Born\,\delta_\obs
+ \int d\sigmah_n^\finvirt\,\delta_\obs+\int d\sigmah_n^{\rm fin.{}\;int}\,\delta_\obs
 +\int\left(d\sigmah_{n+1}^{\rm real}-d\sigmah_{n+1}^{\rm sub}\right) \delta_\obs\,.
\label{eq:NLOfull}
\end{equation}

As in an LO computation, the integrals in \eqn{eq:NLOfull} are most
easily computed by Monte-Carlo sampling.  (In our setup, using
\Sherpa{}~\cite{sherpa1,sherpa2}, the code first adapts an integration grid to the integrand,
and many independent integrations are done using these grids.)  We
generate events separately for each of the four types of
contributions, over $n$-particle phase space for the Born (B), virtual
(V), and integrated-subtraction (I) contributions, and over
$(n+1)$-particle phase space for the subtracted real-emission (R)
contributions.  At lower multiplicities, this split-up is sufficient.
At higher multiplicities, however, a further subdivision of
contributions is desirable.  Different parts of each contribution have
very different computational complexities and magnitudes; for example,
the subleading-color contributions are more costly by at least an
order of magnitude with three or more colored partons, yet give only a
small contribution to cross sections.  A subdivision according to
initial-state parton types is also helpful in this regard.  For each
part, we adapt the number of phase-space points according to its
relative contribution to cross sections.  The subdivision also allows
us to add more statistics to selected parts without having to perform
additional computations for parts whose statistics are adequate.  In
order to keep the subdivision flexible, and make code using them more
robust to future evolution, we simply label the different parts
sequentially within each of the four types, for example
(R001, R002, R003).

As in an LO calculation, we must save all particle four-momenta in
each phase-space configuration in order to be able to recompute (or
compute afresh) observables.  We must also save the renormalization
scale $\mur$ in order to be able to vary it, and the initial-state
parton momentum fractions $x_{1,2}$ and factorization scale $\muf$ in
order to be able to vary the pdfs and thereby estimate pdf
uncertainties.  We must also save each configuration's integration
weight; alongside it, we again save the matrix-element weight as well.

Unlike for an LO calculation, however, these elements do not suffice
in order to be able to vary scales or pdfs.  They do
suffice for the Born and subtracted real-emission contributions, as
both the strong coupling and the pdfs appear as simple overall factors
in the integrand.  However, the virtual contribution contains
additional dependence on $\mur$ arising from the one-loop amplitude;
and the integrated-subtraction contribution contains both additional
dependence on $\mur$ and $\muf$, and different dependence on pdfs for
different parton species.  We must save more detailed information for
these contributions, as we discuss in greater depth in
\sect{ChangingScalesSection}.

The \ntuple{} collections we describe are focused on the computation
of jet cross sections.  The framework we describe is applicable to
other infrared-safe observables, but the specific files we have
generated and are documenting herein rely on a jet algorithm, and can
only be used to compute observables which also are defined using one
of a selected set of jet algorithms.  Modern jet algorithms are
characterized by a clustering (or seedless cone) algorithm; a minimum
jet $\pT$; and a jet size $R$.  The event samples described here allow
for any one of the anti-$\kT$~\cite{AntiKT}, $\kT$\cite{KT}, and
\SISCone{}~\cite{Siscone} algorithms, in each case with jet sizes $R$
chosen from the set $\{0.4,0.5,0.6,0.7\}$.  The \SISCone{} merging
fraction parameter is taken to be $f= 0.75$.  We use the FastJet
library~\cite{FastJet} to implement these jet algorithms.  The minimum
$\pT$ is specified in each file, but is typically $20$~or $25$~GeV.
Only observables imposing this cut, or a tighter one, are allowed for
use with \ntuple{} files.  (If the minimum $\pT$ cut is too tight, the
number of events in the sample passing the cut may be too small for
adequate statistical accuracy.)  These restrictions on jet algorithm
and minimum jet $\pT$ could of course be relaxed by generating new
event samples within our framework.

The main practical tradeoff in this approach is the sizable storage
requirement for the \ntuple{} files. For example, the provided 7 TeV
\ntuple{} files for $W^++3$ jets all together require 50 GB of
storage, while the corresponding files for $W^++4$ jets would require 375
GB. Of course, the
storage requirements depend greatly on the process and on the desired
statistics.  The files are quite
voluminous, especially at higher multiplicities; the compression
offered by \Root{} yields a significant reduction compared to a naive
binary format both in disk-space usage and in transmission times. 

\section{The \nTuple{} Files}
\label{nTupleFileSection}

%%%%%%%%%%%%%%  TABLE %%%%%%%%%%%%%%%%%%%%%
\begin{table}
\tabcolsep=2pt
\begin{tabular}{|c|c|p{8.45cm}|}
\hline
Branch name & Type & Notes \\
\hline
\varref{id} & I & ID of the event. Real-emission entries and their associated counterterms share the same ID. 
\\
\varref{nparticle} & I & number of particles  in the final state \\
\varref{px}, \varref{py}, \varref{pz} & F[\varref{nparticle}] & array of $p_x$, $p_y$, $p_z$ respectively, for final-state particles \\  
\varref{E} & F[\varref{nparticle}] & array of energies $E$ for final-state particles  \\  
\varref{kf} & I[\varref{nparticle}] & PDG codes of the final-state particles \\
\varref{weight} & D & total weight of the entry \\
\varref{weight2} & D & secondary or correlated weight used to compute the subtracted real-emission's statistical errors.  Identical to \varref{weight} for the B, V, and I contributions; the normalization differs for the R contribution\\
\varref{me\_wgt} & D & coefficient of the product of parton-distribution functions in \varref{weight}.  For the B, V,
and R contributions, this is the squared matrix element multiplied by the phase-space measure and the Jacobian from 
\Sherpa's phase-space mapping\\
\varref{me\_wgt2} & D & coefficient of the product of parton-distribution functions in \varref{weight2}\\
\varref{x1}, \varref{x2} & D & 
 fraction of hadron momentum carried by the first and second incoming partons,
respectively \\
\varref{x1p},\varref{x2p} & D & 
 secondary momentum fractions $x_{1,2}'$ used in integrated subtraction 
entries~\cite{AmegicII} \\
\varref{id1}, \varref{id2} & I & PDG codes of the first and 
second incoming partons respectively \\
\varref{fac\_scale} & D & factorization scale used ($\mufz$) \\
\varref{ren\_scale} & D & renormalization scale used ($\murz$) \\
\varref{nuwgt} & I & number of additional weights \\
\varref{usr\_wgts} & D[\varref{nuwgt}] & additional weights needed to recompute the entry's weight for a different scale or pdf choices\\
\varref{part} & C & type of contribution: B, V, I, or R\\
\varref{alphas\_power}  & S & power of the coupling\\
\varref{alphas} & D & $\alpha_s$ value used for this entry \\
\hline
\end{tabular} 
\caption{Branches in a \BlackHat{}+\Sherpa{} \Root{} file. The type of the
  data entry follows \Root{}'s notation. In the second column,
 ``D'' stands for ``double-precision
  floating point number'', ``F'' for ``single-precision floating point
  number'', ``I'' for ``integer'', ``S'' for ``short integer'', and ``C'' for 
  ``character array''. Square brackets denote an array.}
\label{table:BranchList}
\end{table}
%%%%%%%%%%%%%%%%%%%%%%%%%%%%%%%%%%%%%%%%%%%%%%%%%%%%%%

As described in the previous section, we save collections of
phase-space configurations along with additional information.  
Each phase-space configuration represents one event at LO.  
The same is true
for three of the four types of contributions at NLO: the Born (B), the
virtual (V), and the integrated-subtraction (I) contributions.  In the
fourth type of contribution, the subtracted real-emission (R) one,
each event in general will contain multiple configurations: one
corresponding to an emission, with the rest of the configurations
corresponding to subtractions; we will call each of the configurations
`entries'.  Each \Root{} \ntuple{} file contains a collection of
events.  Each (numbered) part within each of the contribution types
(B,V,I,R at NLO) may in general be split up into a number of different
files for convenience.  The event files we describe in this section
were produced using \BlackHat{}~\cite{BH} and \Sherpa{}
\cite{sherpa1,sherpa2,AmegicI,AmegicII,Comix}, according to the setups
described in refs.~\cite{W3PRL,W3,Z3,W4,Z4,Zgamma,pureQCD}.  Each file
is a \Root{} file, containing a set of events, along with information
about the file content, and sample histograms which can be used to
cross-check analyses using the file.  (For a review of \Root{} and its
file formats, the reader may consult ref.~\cite{ROOT}.)
  
Information for the entries in each file is stored in a \Root{} tree
called \varref{BHSntuples}.  The branches of this tree are listed in
Table~\ref{table:BranchList}.  We have chosen to restrict the
numerical precision for the momenta to single precision in order to
limit disk space usage.  This means that the weights cannot be
recomputed exactly from the momenta for each entry; but there is of
course no need to do so.  For this reason, because each individual
weight is in any case not necessarily accurate even to single
precision (see refs.~\cite{BH,W3,Z3,W4jstability,W5j} for examples of
numerical uncertainty distributions of matrix elements), and because
the weights include phase-space and Jacobian factors arising from \Sherpa{}'s
integration grids, they are not suitable as reference points for
verifying or comparing matrix elements.  For such purposes, reference
points and matrix-element values quoted in refs.~\cite{W3,Z3} should
be used.

\subsection{Coordinates and units}

We take the coordinate axes such that the beams are directed along the
$z$ axis, with the initial-state parton with momentum fraction
\varref{x1} and Particle Data Group (PDG) code \varref{id1} moving in
the positive $z$ direction.  The transverse directions are labeled by
$x$ and $y$, with $(x,y,z)$ forming a right-handed coordinate system.
The energies of particles are denoted by $E$ and the spatial momenta
by $(p_x,p_y,p_z)$.  We use natural units ($\hbar=c=1$) and GeV for
energies, masses and momenta.  Weights are normalized to yield cross
sections in picobarns.

\def\optbar{\raisebox{-4.0pt}{\scalebox{.3}{
\textbf{(}}}\raisebox{-3.0pt}{{\_}}\hskip 1pt\raisebox{-4.0pt}{\scalebox{.3}{\textbf{)}}}}
%%%%%%%%%%%%%%%%%%%%%%  TABLE %%%%%%%%%%%%%%%%%%%%

\begin{table}
\begin{tabular}{|c|p{2.28 in}|l|}
\hline
Process & \ntuple{} file sets & References \\
\hline
$W^{\pm}(\rightarrow e^\pm\overset{\optbar}{\nu})+0,1,2\mbox{ jets }$ & B001, I001, R001, V001
&\cite{W3PRL,W3}\\
$W^{\pm}(\rightarrow e^\pm\overset{\optbar}{\nu})+3\mbox{ jets }$ & B001, I001, R001, V001--V002
&\cite{W3PRL,W3}\\
$W^{-}(\rightarrow e^-\bar\nu)+4\mbox{ jets }$ & B001, I001, R001, V001
&\cite{W4, ItaOzeren}\\
$W^{+}(\rightarrow e^+\nu)+4\mbox{ jets }$ & B001, I001, R001--R005, V001
&\cite{W4, ItaOzeren}\\
$Z(\rightarrow e^+e^-)+0,1,2\mbox{ jets }$ & B001, I001, R001, V001
&\cite{Z3}\\
$Z(\rightarrow e^+e^-)+3\mbox{ jets }$ & B001, I001, R001, V001--V002
&\cite{Z3}\\
$Z(\rightarrow e^+e^-)+4\mbox{ jets }$ & B001, I001--I003, R001--R006, V001--V006
&\cite{Z4}\\
$n\mbox{ jets }$ ($n=1,2,3,4$) & B001, I001, R001, V001
&\cite{pureQCD}\\
\hline
\end{tabular} 
\caption{Available processes at NLO, and their decomposition into 
\ntuple{} file sets.}
\label{table:availableProcesses}
\end{table}
%%%%%%%%%%%%%%%%%%%%%%%%%%%%%%%%%%%%%%%%%%%%%

\subsection{\nTuple{} Collections}

In this article, 
we document \ntuple{} files for the LHC processes listed in Table
\ref{table:availableProcesses}, which we 
have made available as described in \sect{LocationSection}.
The corresponding references for each class of processes are
given in the last column.

As explained in section~\ref{NLOCalculationsSection},
the events for each process are split up into
different types according to the different
terms in \eqn{eq:NLOfull}. They are further subdivided into
different parts, organized in
subdirectories, with each part in turn split into multiple files. 
We must add together contributions from all parts within all types
of contributions to obtain the complete NLO cross sections or
distributions, as we describe in greater detail in 
sections~\ref{NTupleUseSection} and~\ref{ChangingScalesSection}.
 The different parts are independent, analogous to 
different subprocesses, so their statistical integration errors can be
added in quadrature.  We show the list of the different parts 
for each process in \Tab{table:availableProcesses}.

The \ntuples{} were generated using the MSTW2008~\cite{MSTW2008} NLO
parton distribution functions, and a five-flavor running
$\alpha_s(\mu)$ where the value of $\alpha_s(M_Z)$ is specified by the
parton distribution set.  The renormalization and factorization scales
are,
\begin{equation}
\mufz = \murz
= {\textstyle\frac12}\HTpartonicp \equiv 
\mbox{\footnotesize $\displaystyle \frac12$}\Big(\sum_j \pT^j + \ET^V\Big)\,.
\label{HTpartonicp}
\end{equation}
In this equation, originally given in ref.~\cite{W4},
 the sum runs over all partons in the final state, with
$\pT^j$ the transverse momentum of the $j$th parton, and $\ET^V$ is
the transverse energy of the vector boson,
\begin{equation}
\ET^V = \sqrt{m_V^2 + (\pT^V)^2}\,.
\end{equation}
(For pure-jet processes, the $\ET^V$ term is of course omitted.)
As we explain in more detail in \sect{ChangingScalesSection},
the \ntuples{} can be used to generate predictions for other
parton distributions, and other choices of renormalization
and factorization scales\footnote{Cross sections and distributions
computed using the \ntuples{} documented here may not match the
results 
quoted in the references in the last column of \Tab{table:availableProcesses}
exactly, because the earlier results used six-flavor running (above the top-quark mass) and
a different definition of the leading-color approximation.}.

The {\tt getInfo} program (see section \ref{sec:getInfo})
provides useful information about the process, input parameters,
cuts and jet algorithms for a specific \ntuple{} file.

For processes containing a single electroweak boson,
we fold in its decay into a massless lepton pair.
We label the pair by $e^+\nu$ for a $W^+$ boson, $e^-\bar{\nu}$ 
for a $W^-$ boson, and $e^+e^-$ for a $Z$ boson.  Results for
the corresponding 
muon channels are identical in the massless-lepton approximation.
Off-shell effects are taken into account by distributing the lepton-pair
invariant mass in a relativistic Breit-Wigner profile with width $\Gamma_V$
around the mass $m_V$ of the electroweak vector boson $V$.
In the case of the $Z$ boson (that is, $e^+e^-$ final states), we include
virtual photon $\gamma^*$ exchange as well.  
For all processes we neglect
the contribution of a massive top quark in the quark loops.

\subsubsection{\Wjn-jet processes}

For \Wjn-jet processes, we take the CKM matrix to be diagonal.
We provide separate sets of files for \Wmjn-jet and \Wpjn-jet production.
The files we provide yield results for the \Wjjjj-jet process 
in the leading-color approximation for the virtual contributions.
The \ntuple{} collections impose no cuts
 on the generated momenta of the final-state
electron or neutrino, no isolation cuts between the
leptons and the jets, and no cuts on
the jet rapidities.  We impose cuts only on the transverse momenta of the jets,
\begin{equation}
\begin{aligned}
\pT^{\rm jet} &> 25 \mbox{ GeV}\,,\qquad 7~\mbox{TeV~\ntuple{}s~for~\Wjjjjx~jets}\,,\\
\pT^{\rm jet} &> 20 \mbox{ GeV}\,,\qquad 8~\mbox{TeV~\ntuple{}s~for~\Wjjjx~jets}\,.
\end{aligned}
\end{equation}
As noted in \sect{NLOCalculationsSection}, the allowed jet algorithms are
the anti-$\kT$, $\kT$, and \SISCone{} ones, in each case with jet sizes
$R$ chosen from the set $\{0.4,0.5,0.6,0.7\}$.  In the \SISCone{} case
the merging parameter is taken to be $f= 0.75$. 

%%%%%%%%%%%%%%%%%%%%
\subsubsection{\Zgamjn-jet processes}

In contrast to the $W\,+\,n$-jet processes, for $Z$ processes we also
impose a cut on the invariant mass of the $e^+e^-$ pair in a window
around the $Z$ mass, 
$60 < M_{e^+ e^-} < 120  \mbox{ GeV}$,
in order to suppress the contribution of the
virtual photon.  Otherwise, we again impose only a cut on the
jet transverse momenta,
\begin{equation}
\begin{aligned}
\pT^{\rm jet} &> 25 \mbox{ GeV}\,,\qquad 7~\mbox{TeV~\ntuple{}s~for~\Zgamjx~jets}\,,\\
\pT^{\rm jet} &> 20 \mbox{ GeV}\,,\qquad 7~\mbox{TeV~\ntuple{}s~for~\Zgamjjjx~jets}\,,\\
\pT^{\rm jet} &> 20 \mbox{ GeV}\,,\qquad 8~\mbox{TeV~\ntuple{}s~for~\Zgamjjjxx~jets}\,.
\end{aligned}
\end{equation}
The allowed jet algorithms are the same as given above for \Wjn-jet processes.

The virtual contributions for the $Z\,+\,4$-jet process are again
computed in the leading-color approximation described in
ref.~\cite{Z4}.  Based on studies of lower multiplicities~\cite{W3,Z3} and
also in the case of $W+4$-jets~\cite{ItaOzeren}, we expect these
neglected pieces to be on the order of 3\% of the total cross section.  As in
ref.~\cite{Z4}, we also drop the axial- and vector-coupling loop
contributions, along with the effects of top quarks in the loop.  We
expect these neglected pieces to contribute under 1\%.
If we neglect the small effect from the muon mass, these
\ntuples{} are just as valid for the $Z$ boson decaying into
a pair of muons.

%%%%%%%%%%%%%%%%%%%%
\subsubsection{Pure-jet processes}

For pure-jet processes, we impose the cut,
\begin{equation}
\pT^{\rm jet} > 40 \mbox{ GeV}\,.
\end{equation}
We treat the five light-flavor quarks as massless, and drop
top-quark loops, following
ref.~\cite{pureQCD} (this has a sub-percent effect). 
We include the full color dependence of all
contributions. 

\subsection{Location}
\label{LocationSection}

An up-to-date list of available processes is maintained at

{\tt http://blackhat.hepforge.org/trac/wiki/Availability}
 
The locations from which the \ntuple{} files may be obtained are given in
{\tt http://blackhat.hepforge.org/trac/wiki/Location}.  They are currently available
at CERN on CASTOR and on the LHC Grid.  At each location, the files are in\hfil\break 
\null\hspace*{5mm}{\tt $\langle${\rm base}$\rangle$/BHSNtuples/PROCESS/ENERGY/PART}\hfil\break
where {\tt ENERGY\/} is either {\tt 7TeV} or {\tt 8TeV},
and {\tt PART\/} is one of {\tt B\/}, {\tt V\/}, {\tt I\/}, or {\tt R\/}.
\subsection{Checks}\label{sec:checks}

All \ntuple{} files contain several histograms which can be used to
check their consistency and the implementation of the program reading them. 
The histograms are listed in Table~\ref{tab:histograms}.

%%%%%%%%%%%%%%%%%% TABLE %%%%%%%%%%%%%%%%%
\begin{table}
\begin{tabular}{|c|c|c|}
\hline
\varref{xsection} & total cross section & 1 bin of width $1$\\
\varref{h\_pt\_jN} & transverse momentum of jet N & $50$ bins: $[0,1000]$\\
\varref{h\_eta\_jN} & pseudo-rapidity of jet N & $22$ bins: $[-4.4,4.4]$\\
\varref{h\_pt\_NP} & trans.{} momentum of the non-parton NP & $200$ bins $[0,2000]$ \\
\hline
\end{tabular}
\caption{List of the histograms provided to test validate the analysis
of an \ntuple{} file. A non-parton can be an electron, positron, photon,
neutrino or anti-neutrino
\label{tab:histograms}
}
\end{table}
%%%%%%%%%%%%%%%%%%%%%%%%%

Each of the histograms listed in Table~\ref{tab:histograms} comes in
two copies, one in which the distributions are computed using the
original weights, without change of renormalization or factorization scale,
and a second in which both $\mur$ and $\muf$ have been changed to 
$\HT^{\rm all}$ , defined to be the scalar sum
of the transverse momentum of all particles (including the neutrino) in
the event\footnote{This scale is not appropriate for
observables sensitive to the polarization of the vector boson, as
explained in ref.~\cite{Wpol}.}. The histogram names have
{\tt \_Orig} and {\tt \_HTallp} respectively appended to them.

The program {\tt getInfo} (see section \ref{sec:getInfo}) can be used to obtain
the cuts and jet algorithm used for the histograms.

\section{Using the \ntuple{} files}
\label{NTupleUseSection}

One can analyze the entries in the \ntuple{} files to generate histograms with a wide variety
of experimental cuts, so
long as all of the following conditions are met:
\begin{itemize}
\item One of the jet algorithms used in generating the \ntuple{} file
  is applied.  (The list of compatible jet algorithms and parameters
  are stored in each file but are the same for all files for the same
  process and center-of-mass energy.  See section~\ref{sec:getInfo}
  for a description of this information and how to extract it from an
  \ntuple{} file.)
\item The number of jets passing the algorithm and associated cuts is
  at least $n$ for an $n$-jet process.
\item All cuts and observables are defined in terms of jets passing the cuts and not in
  terms of partons.  (It is perfectly acceptable, and indeed often desirable, to define the renormalization
  and/or factorization scales using parton momenta so long as the definition
  is infrared- and collinear-safe.)
\end{itemize}

As explained above, each type of contribution (`B', `R', $\ldots$) to a process is split up
into a number of parts (R001, R002, $\ldots$).  The number of parts will depend on the
process and type of contribution.  An analysis should in general sum
over all available parts and all types of contributions.
(Once one has verified that a given part yields a negligible 
contribution to the observables of interest compared to the desired accuracy,
one may choose to omit it.)  The events for each part may be split up into a
number of files; each file will contain an independent sample of
events.  An analysis should sum over as many files as required to
obtain the desired statistical accuracy for the given part and type of
contribution, but need not sum over all available files.  Summing over
more files for a given part will increase the statistical accuracy of
estimates for that part.

\def\cuts{{\rm cuts}}
To compute the contribution of a given type $t$ and part $p$ to an observable,
one must sum the weight times the observable's value over all entries in all chosen
files, and normalize by the total number of entries in these files,
\begin{equation}
\langle\Obs\rangle^{(t,p)} = \frac1{N_{t,p}}\sum_{e=1}^{N_{t,p}} w_{t,p,e} \Obs_{t,p,e}\,,
\label{ObservablePart}
\end{equation}
where $N_{t,p}$ is the total number of entries (across all files) for the given part of the
given type, $w_{t,p,e}$ the weight of the entry (given by the
\varref{weight} branch of the \Root{} file), and $\Obs_{t,p,e}$ the
value of the observable evaluated on the entry.  The value of an
observable is then giving by summing over all parts of all types of
contributions,
\begin{equation}
\langle\Obs\rangle = \sum_{t\in T,p\in P_t} \langle\Obs\rangle^{(t,p)}\,.
\label{Observable}
\end{equation}
where $T$ is the set of types, and $P_t$ the set of parts in type $t$.  As 
a simple example, if we choose $\Obs_{t,p,e} = \Theta_{\cuts}$
(where $\Theta_{\cuts} = 1$ if the entry is allowed by our chosen jet algorithm
and all other applied analysis cuts, and $0$ otherwise),
 we obtain the
total cross section.  In \eqn{ObservablePart}, all types of
contributions are computed in a similar manner.
The weight itself can be recomputed to alter the factorization scale, the renormalization
scale, or the pdf, as we shall explain in \sect{ChangingScalesSection}.
The overwhelming fraction of the computation effort in an analysis is in 
the execution of \eqn{ObservablePart}.  One can combine different batches or
sets of files part-by-part by weighting with the total number of entries in each,
\begin{equation}
\langle\Obs\rangle^{(t,p)}_{1+2} = \frac1{N_{t,p,1+2}} \big(
N_{t,p,1}\langle\Obs\rangle^{(t,p)}_1+N_{t,p,2}\langle\Obs\rangle^{(t,p)}_2\big)
\label{CombiningBatchesObservable}
\end{equation}
(where $N_{t,p,1+2} = N_{t,p,1}+N_{t,p,2}$),
and then re-evaluating \eqn{Observable}.
One would of course typically compute several observables in parallel during a single pass over
the selected files. 

\def\Nt{{\widetilde N}}
\def\Nh{{\widehat N}}
\def\RS{\textrm{R}}
\def\wt{{\widetilde w}}
\def\error{\varepsilon}

In order to compute the error estimate for the observable, we must
however treat the subtracted real-emission contributions (R)
specially.  For each part in each of the other types of contributions
(B, V, and I), the error estimate is the standard one for Monte-Carlo
integration,
\begin{equation}
\error_\Obs^{(t,p)} = \frac1{\sqrt{N_{t,p}(N_{t,p}-1)}}
 \biggl[ \sum_{e=1}^{N_{t,p}} \bigl(w_{t,p,e}\Obs_{t,p,e}\bigr)^2 
  - \frac1{N_{t,p}} \biggl(\sum_{e=1}^{N_{t,p}} w_{t,p,e} 
                                         \Obs_{t,p,e}\biggr)^2\biggr]^{1/2}\,.
\label{ErrorEstimatePart}
\end{equation}
(We denote the error estimate by $\error$ in order to avoid confusion
with the cross section $\sigma$.)
The R contributions are special, because each event in general
contains more than one entry.  (In the \Root{} file, all entries within a given
event share the same \varref{id}.) The entries (or phase-space
configurations) are of two different kinds, a real-emission
configuration and counter-configurations.  The latter correspond to
the subtraction term that regulates the squared matrix element in the
soft and collinear limits, where the unregulated matrix element
diverges.  The real-emission configurations and subtraction
counter-configurations for a given event are strongly anticorrelated.
This means that simply adding the weights independently as in
\eqn{ErrorEstimatePart} will grossly overestimate the statistical
error.  (The estimated error would typically be of order the central value.
Using these weights will, however, yield the correct central value in
\eqn{ObservablePart}.)  The anticorrelation must be taken into account
properly, which we can do using the following formula,
\begin{equation}
\begin{array}{r@{\hskip 0pt}l}
\displaystyle
\error_{\Obs}^{(\RS,p)} = \frac1{\sqrt{\Nt_{\RS,p}(\Nt_{\RS,p}-1)}} \biggl[&
\displaystyle\hskip 1mm
  \sum_{e=1}^{\Nt_{\RS,p}} \biggl(\sum_{j=1}^{\Nh_{p,e}} \wt_{\RS,p,e,j}
                                                   \Obs_{\RS,p,e,j} \biggr)^2
\\
&\displaystyle\hskip 1mm
- \frac1{\Nt_{\RS,p}} 
    \biggl(\sum_{e=1}^{\Nt_{\RS,p}} \sum_{j=1}^{\Nh_{p,e}} \wt_{\RS,p,e,j}
                                                   \Obs_{\RS,p,e,j}\biggr)^2
\biggr]^{1/2}\,,
\end{array}
\label{ErrorEstimateRS}
\end{equation}
where $\Nt_{\RS,p}$ is the total number of {\sl events\/} (or
equivalently, different (filename,ID) pairs with the ID given by the \varref{id} branch) 
in the files analyzed for the given part
$p$, in contrast to the total number of {\sl entries\/} $N_{\RS,p}$;
where $\Nh_{p,e}$ is the number of different entries in the given
event, and where $\wt_{\RS,p,e,j}$ is the secondary or correlated
weight of the entry (given by the \varref{weight2} branch of the
\Root{} file).  For simplicity of use, events are ordered in the
\ntuple{} files, with real-emission entries and counter-entries with
the same \varref{id} appearing contiguously.

If we choose to combine different batches of files using \eqn{CombiningBatchesObservable},
we also need the corresponding formula for combining uncertainty estimates,
\begin{equation}
\begin{aligned}
\error_{\Obs,1+2}^{(t,p)} = &\frac1{\sqrt{N_{t,p,1+2}(N_{t,p,1+2}-1)}}\\
&\times\bigg[ N_{t,p,1}(N_{t,p,1}-1) \big(\error_{\Obs,1}^{(t,p)}\big)^2
         +N_{t,p,2}(N_{t,p,2}-1) \big(\error_{\Obs,2}^{(t,p)}\big)^2\bigg]^{1/2}\,,
\end{aligned}
\end{equation}
for the B, V, and I contributions, and a similar equation with 
$N_{t,p}\rightarrow\Nt_{{\rm R},p}$ for the
R contribution. 

The different types of contributions are statistically independent, as
they are generated using independent sets of phase-space
configurations, and so the overall error estimate simply adds the
error estimates for the different parts in quadrature,
\begin{equation}
\error_\Obs = \Bigl[
 \sum_{t\in T,p\in P_t} \bigl(\error_\Obs^{(t,p)}\bigr)^2\Bigr]^{1/2}\,.
\end{equation}

\def\dh{\hat\delta}
In addition to overall values of observables, we will typically want to
obtain values of distributions as well.  To obtain the 
prediction for the distribution in a
variable $v$, we must divide its range $B$ into bins (possibly including
overflow and underflow bins).  For each bin, the expected value is given
by,
\begin{equation}
\left\langle \frac{d\sigma}{dv}\right\rangle_b  =
\sum_{t\in T,p\in P_t} \left\langle\frac{d\sigma}{dv}\right\rangle^{(t,p)}_b\,,
\end{equation}
where
\begin{equation}
\left\langle\frac{d\sigma}{dv}\right\rangle^{(t,p)}_b
= \frac{1}{\Delta_b}\frac1{N_{t,p}}\sum_{e=1}^{N_{t,p}} w_{t,p,e} 
\dh(v_{t,p,e},b)\,,
\label{DistributionPart}
\end{equation}
where $\Delta_b$ is the bin's width (in the units of the observable), 
and where $\dh(v,b)$ is 1 if the value lies in the given bin $b$,
and 0 otherwise.
The values of the distribution 
for different bins can of course be obtained in parallel,
by assigning each file entry to the appropriate bin.  This formula
is what is typically implemented by histogramming codes, which can
be used straightforwardly to process the \ntuple{} files for all the types of
contributions.

For the error estimates, we must again treat the subtracted real-emission
contribution specially.  For each part in each of the other types
(B, V, and I), we have a standard form for the error estimate,
\begin{equation}
\begin{array}{r@{\hskip 0pt}l}
\displaystyle
\error_{d\sigma/dv,b}^{(t,p)} = \frac1{\Delta_b}\frac1{\sqrt{N_{t,p}(N_{t,p}-1)}}
\biggl[&\displaystyle\hskip 1mm
 \sum_{e=1}^{N_{t,p}} w_{t,p,e}^2 \dh(v_{t,p,e},b)\\
&\displaystyle\hskip 1mm \null
- \frac1{N_{t,p}} \biggl(\sum_{e=1}^{N_{t,p}} w_{t,p,e} 
                                    \dh(v_{t,p,e},b)\biggr)^2\biggr]^{1/2}\,.
\end{array}
\label{DistributionErrorEstimatePart}
\end{equation}
This again is the formula that will
typically be the one implemented by histogramming
routines.

For the R contribution, we must take the anticorrelation into account,
so that the error estimate is given by,
\begin{equation}
\begin{array}{r@{\hskip 0pt}l}
\displaystyle
\error_{\Obs}^{(\RS,p)} = \frac{1}{\Delta_b}\frac1{\sqrt{\Nt_{\RS,p}(\Nt_{\RS,p}-1)}} \biggl[&
\displaystyle\hskip 1mm
  \sum_{e=1}^{\Nt_{\RS,p}} \biggl(\sum_{j=1}^{\Nh_{p,e}} \wt_{\RS,p,e,j}
                                                 \dh(v_{t,p,e,j},b) \biggr)^2\\
&\displaystyle\hskip 1mm \null
-  \frac1{\Nt_{\RS,p}} 
   \biggl(\sum_{e=1}^{\Nt_{\RS,p}} \sum_{j=1}^{\Nh_{p,e}} \wt_{\RS,p,e,j}
                                                   \dh(v_{t,p,e,j},b)\biggr)^2
\biggr]^{1/2}\,.
\end{array}
\label{DistributionErrorEstimateRS}
\end{equation}
 A given event is counted as being in a bin if any of its entries are in a bin;
a given event can therefore appear in more than one bin. The extent to which this happens,
and hence the error estimates, will
depend on the value of the integrated subtraction's 
$\alpha_{\rm dipole}$ parameter~\cite{Nagy}, 
which is $0.03$ for all the \ntuple{} files documented here.
This formula is {\it not\/} what is implemented by standard
histogramming routines; in particular, current versions of 
R{\sc ivet\/}~\cite{Rivet} do not take
anticorrelations into account, and so will not compute statistical
error estimates correctly for this contribution.  The user will
typically need to supply his or her own histogramming routines to
obtain correct estimates.

Histograms with statistical uncertainty estimates
obtained using \eqn{DistributionErrorEstimateRS} cannot be rebinned in the usual way (that is keeping track of the sum of the weights and the sum of the weights squared for each bin). To understand why, let us consider two neighboring
bins $b_1$ and $b_2$ and see what happens if we want to compute the error associated with the combined bins which 
we will refer to as $b_1+b_2$. The second term in the calculation of the error will be the same, whether one considers 
the bins separately or not. Let us consider the first term,
\begin{eqnarray}
\lefteqn{\sum_{e=1}^{\Nt_{\RS,p}} \biggl(\sum_{j=1}^{\Nh_{p,e}} \wt_{\RS,p,e,j}
                                                 \dh(v_{t,p,e,j},b_1+b_2) \biggr)^2}&&\nonumber\\
&=&\sum_{e=1}^{\Nt_{\RS,p}} \biggl(\sum_{j=1}^{\Nh_{p,e}} \wt_{\RS,p,e,j}
                                                 \left(\dh(v_{t,p,e,j},b_1)+\dh(v_{t,p,e,j},b_2)\right) \biggr)^2 \nonumber\\
&=&\sum_{e=1}^{\Nt_{\RS,p}} \biggl[
\biggl(\sum_{j=1}^{\Nh_{p,e}} \wt_{\RS,p,e,j}\dh(v_{t,p,e,j},b_1) \biggr)^2 +
\biggl(\sum_{j=1}^{\Nh_{p,e}} \wt_{\RS,p,e,j}\dh(v_{t,p,e,j},b_2) \biggr)^2 
 \nonumber\\
&&+
2\biggl(\sum_{j,j'=1;j\neq j'}^{\Nh_{p,e}} \wt_{\RS,p,e,j}\dh(v_{t,p,e,j},b_1)\wt_{\RS,p,e,j'}\dh(v_{t,p,e,j'},b_2) \biggr) 
\biggr]
\label{rebinningError}
\end{eqnarray}
The terms on the penultimate line are typically recorded (as the
statistical uncertainty estimate) for each bin alongside the sum of
weights. The term on the last line is not. In the absence of
correlations between entries in an event --- if $\Nh_{p,e}=1$ --- the
double sum over $j$ and $j'$ would disappear, and we would recover the
usual formula for combining uncertainty estimates of two bins.
Because the weights of real-emission and counter-configuration entries
in an event are typically anticorrelated, however, the last term
typically will be negative and will reduce the uncertainty estimate
significantly.  Accordingly, naively rebinning histograms with
uncertainties computed according to \eqn{DistributionErrorEstimateRS}
but then combined in quadrature will overestimate the statistical
uncertainty. The same argument applies to extracting cumulative
distributions from a histogram. If the user plans to rebin or extract
cumulative distributions, the last term in \eqn{rebinningError} should
be recorded for all bin pairs along with the sum of the squared
weights.

We again obtain the overall error estimate for each bin
by adding the separate contributions in quadrature,
\begin{equation}
\error_{d\sigma/dv,b} = \Bigl[
 \sum_{t\in T,p\in P_t} \bigl(\error_{d\sigma/dv,b}^{(t,p)}\bigr)^2\Bigr]^{1/2}\,.
\end{equation}

\section{Changing Scales and Parton Distributions}
\label{ChangingScalesSection}

The \BlackHat{}+\Sherpa{} \ntuple{} files
contain the information needed to recompute
the weight of the event for a different
renormalization or factorization scale, or for
a different pdf set.  The new weights can be used to compute new
central values, or to compute scale-variation bands and pdf-uncertainty
estimates. In the following subsections, we explain the
additional information stored in the \ntuple{} files, and how
to make use of it.  We denote by $f_1$ and $f_2$ the pdf of the
first and second hadron respectively.
  For the LHC, in which both beams are protons, 
they both correspond to proton pdfs.  \varref{Variables}
shown in a distinct font correspond to branches in 
the \Root{} file. Indices in arrays such as
\varref{usr\_wgts} are zero-based.

\subsection{Born and subtracted-real contributions}

The new weight is given by
\begin{eqnarray}
n &=& \varref{alphas\_power}\,,\\
w &=& \varref{me\_wgt2}\, f_1(\varref{id1},\varref{x1},\muf)
 \, f_2(\varref{id2},\varref{x2},\muf)
 \ \frac{\alpha_s(\mur)^n}{(\varref{alphas})^n} \,,
\end{eqnarray} 
where $\muf$ is the new factorization scale, $\mur$ the new
renormalization scale, $f_{1,2}$ the new pdf, $\alpha_s$ the
corresponding running coupling, and $n$ the power of the strong
coupling $\alpha_s$.  (In the case of $W$ or $Z\,+\,n_j$-jet
processes, this power is $n_j$ in the Born (B) contribution and
$n_j+1$ in the subtracted-real (R) contribution.)  The new scales must
be infrared- and collinear-safe functions of the final-state momenta.
If the factorization scale and pdf set are left unchanged, one can
simplify the computation of $w$, eliminating the pdf function call:
\begin{eqnarray}
w &=& \varref{weight2}\ \frac{\alpha_s(\mur)^n}{(\varref{alphas})^n} \,.
\end{eqnarray}
Alternatively, if we leave the renormalization scale unaltered, 
we could simplify the
computation of $w$ to:
 \begin{eqnarray}
w &=& \varref{me\_wgt2} \, f_1(\varref{id1},\varref{x1},\muf)
  \, f_2(\varref{id2},\varref{x2},\muf) \,.
\end{eqnarray}
For the Born contribution (but not the subtracted-real one), \varref{weight} 
and \varref{weight2} are the same.

\subsection{Virtual contribution}
\label{VirtualContributionSection}

\def\vL#1{\mathcal{L}^{(#1)}}
The virtual contribution (V) is treated in a similar way as the real and 
Born contributions, except that the matrix element has an explicit dependence
on the renormalization scale.  In dimensional regularization, this dependence
arises from the introduction of a scale to give the coupling $g$ the required
dimension, $g\rightarrow g \mu^\epsilon$, along with the $\overline{\rm MS}$ ultraviolet subtraction
that replaces the bare coupling $g_0$ with the physical coupling $g(\mur)$,
\begin{equation}
g^n \mur^\epsilon c_\Gamma {\cal \hat A}_n^{(1)} \longrightarrow
g^n \mur^\epsilon c_\Gamma {\cal \hat A}_n^{(1)}
-b_0 (n-2)g^n c_\Gamma {\cal \hat A}_n^{(0)}\,,
\end{equation}
where $b_0 = 11/2-n_f/3$, $c_\Gamma =
\Gamma(1+\epsilon)\Gamma^2(1-\epsilon)/
\big((4\pi)^{2-\epsilon}\Gamma(1-2\epsilon)\big)$,
and ${\cal \hat A}^{(0,1)}$ are respectively the tree-level and
one-loop amplitude with factors of $g$ and $c_\Gamma$ removed.  This
gives rise to a term of the form $b_0 \alpha_s {\cal A}^{(0)}
\ln(\mur/s)$.  Because $\mur$ also enters terms with infrared
divergences, there are also double-logarithmic terms in the virtual
contribution. In the sum over all contributions, they will cancel
statistically
against similar terms in the integrated-subtraction term, leaving only
double logs of ratios of invariants as well as the linear
term in $\ln\mur$.  A dependence on $\ln\mur$
beyond linear will remain from the running of the strong coupling
$\alpha_s$.
In calculations using \ntuples{}, the dependence on $\murf$
is taken into account using the additional
weights \varref{usr\_wgts}.  These weights are computed from the
coefficients $\vL{1,2}$ of the $1/\epsilon$ and $1/\epsilon^2$
poles respectively in the virtual contribution [with the coupling
removed],
\begin{equation}
\begin{aligned}
\varref{usr\_wgts[0]} &= \frac1{2\pi} \vL1 \Phi\,,\\
\varref{usr\_wgts[1]} &= \frac1{2\pi} \vL2 \Phi\,,
\end{aligned}
\label{usrweightsI}
\end{equation}
and saved in the \ntuple{} file.  The coefficient $\vL1$ depends on
$b_0$.  In this equation, $\Phi$ is the
phase-space measure at the configuration for the event, including
Jacobian factors from any remappings used by the phase-space
generator.  The $b_0$ terms mentioned above reside within $ \vL1$.  An
analysis code can then recompute the event weight using the following
formulas,
\begin{eqnarray}
n &=& \varref{alphas\_power}\,,\\
l &=& \ln\left(\frac{\mur^2}{\varref{ren\_scale}^2}\right) \,,\label{Vlogdef}\\
m &=& \varref{me\_wgt} + l \, \varref{usr\_wgts[0]}
     +\frac{l^2}{2} \, \varref{usr\_wgts[1]} \,,\label{Vmvariation} \\
w &=& m\, f_1(\varref{id1},\varref{x1},\muf)
    f_2(\varref{id2},\varref{x2},\muf) \, \frac{\alpha_s(\mur)^n}
   {(\varref{alphas})^n} \,.
\end{eqnarray}
In these equations, \varref{ren\_scale} is the reference renormalization
scale $\murz$.
If a new pdf set is chosen, the $\alpha_s$ should correspond to that
set.  If we do not change the renormalization scale, the
scale-changing logarithm in \eqn{Vlogdef} will vanish, and along with
it the additional terms in \eqn{Vmvariation}.  (In the case of
$W$ or $Z\,+\,n_j$-jet processes, $n$ is $n_j+1$ for this
contribution.)

\subsection{Integrated subtraction contribution}

\def\vI#1{\mathcal{V}^{(#1)}}
\def\Msq{|\mathcal{M}|^2}
\def\Msqi#1{|\mathcal{M}_{#1}|^2}
The computation of the new weight for the integrated subtraction
contribution (I) is the most complicated.  
The \ntuple{} file has 16
additional weights $\{$\varref{usr\_wgts[2]},$\ldots$,\varref{usr\_wgts[17]}$\}$ which 
make a re-computation possible.  These weights are computed
from the virtual pole coefficients defined in 
sect.~\ref{VirtualContributionSection}, along with other
quantities extracted from
the Catani--Seymour subtraction formalism~\cite{CataniSeymourLong}.
These additional quantities include the 
coefficients
$\vI{1,2}_i$ of the $1/\epsilon$ and $1/\epsilon^2$
poles respectively in the integrated subtraction functions
for parton $i$,
extracted from eqs.~(5.32--5.34)
 of ref.~\cite{CataniSeymourLong},
\begin{align}
  \vI2_i&=\left\{\begin{array}{ll}
    C_F\,, & i=q\,,\\[1em]
    C_A\,, & i=g\,;
  \end{array}\right.\\
  \vI1_i&=\left\{\begin{array}{ll}
    \displaystyle\frac{3}{2}C_F\,, & i=q\,,\\[1em]
    \displaystyle\frac{11}{6}C_A-\frac{n_f}{3}\,, & i=g\,,
\label{V1Def}
  \end{array}\right.  
\end{align}
($C_R$ is the Casimir in representation $R$ of $\textrm{SU}(3)$);
the color- (and helicity-) summed squared tree-level matrix
element with factors of the strong coupling removed,
$\Msq$; a logarithm $\ell_{ik} = \ln \mufz^2/|2p_i\cdot p_k|$;
and the color-correlated squared tree-level matrix element
as given [up to a different normalization] in eq.~(3.13) of 
ref.~\cite{CataniSeymourLong},
\def\Mcc#1{\mathcal{M}_{#1}}
\begin{equation}
\Mcc{ik} = \left\langle 1,\ldots,n\right|
 \frac{\mathbf{T}_i\cdot\mathbf{T}_k}{\mathbf{T}_i^2}
\left|1,\ldots,n\right\rangle\,,
\end{equation}
where $|1,\ldots,n\rangle$ is the tree amplitude in the color--helicity
basis of ref.~\cite{CataniSeymourLong} (again, with factors of the
strong coupling removed).

\def\reg{\mathrm{reg}}
\def\oK{{\overline{\kern -2pt K}\kern 2pt}}
\def\tP{{\breve P}}
\def\tK{{\breve K}}
\def\aP{{\widetilde P}}
\def\Li{\mathop{\rm Li}\nolimits}
We also need the 
four-dimensional regularized Altarelli--Parisi splitting functions
$P^{ab}(x)$, given
in eqs.~(5.85--5.88) of ref.~\cite{CataniSeymourLong};
an additional function, related to the splitting functions,
\begin{equation}
\begin{aligned}
    \tP^{q,q}(x)&=2C_F\Big(\ln\frac{1}{1-x}-x-\frac{x^2}{2}\Big)\,,\\
    \tP^{g,g}(x)&=2C_A\ln\frac{1}{1-x}\,,\\
    \tP^{g,q}(x)&=\tP^{q,g}(x)=0\,;
\end{aligned}
\end{equation}
the auxiliary functions $\oK^{ab}(x)$, defined in eqs.~(8.32--8.25)
of ref.~\cite{CataniSeymourLong}; and additional functions,
\begin{equation}
\begin{aligned}
    \tK^{q,q}(x)&=2C_F\Big(\Li_2(x)-\frac{1}{2}\ln^2(1-x)
               +\ln(x)\ln(1-x)\Big)\\
      &\hphantom{=}
      -C_F\Big(5-\pi^2\Big)\,,\\
    \tK^{g,g}(x)&=2C_A\Big(\Li_2(x)-\frac{1}{2}\ln^2(1-x)+\ln(x)\ln(1-x)\Big)\\
      &\hphantom{=}
      -\Big[\Big(\frac{50}{9}-\pi^2\Big)C_A-\frac{8}{9}n_f\Big]\,,\\
    \tK^{q,g}(x)&=\tK^{g,q}(x)=0\,,
\end{aligned}
\end{equation}
and
\begin{equation}
\aP_i(x) = \gamma_i \big(1-\ln (1-x)\big)\,,
\end{equation}
where $\gamma_i = {\cal V}_i^{(1)}$ is the collinear anomalous dimension
given in eq.~(5.90) of ref.~\cite{CataniSeymourLong}, with  
${\cal V}_i^{(1)}$ defined in \eqn{V1Def}.

In terms of these quantities, we have the following expressions for
the set of \varref{usr\_wgts},
\begin{equation}
\null\hskip -4mm\begin{alignedat}{2}
\varref{usr\_wgts[0]}  &= \Msq \Phi
  \Big[\vL1+n_B\,b_0\,
    -\sum_{i\ne k} \frac{\mathcal{M}_{ik}}{\Msq}
     \Big(\vI1_i
     +\vI2_i \ln \frac{\murz^2}{|2p_i\cdot p_k|}\Big)\Big]\,,\\
\varref{usr\_wgts[1]} &=  \Msq \Phi
  \Big[\vL2-\sum_{i\ne k} \frac{\Mcc{ik}}{\Msq}
    \vI2_i\Big]\,,\\
\varref{usr\_wgts[2]} &= \Msq\Phi
  \Big[ \tK^{q,a}(x)+\Big( \delta^{qa}\sum_{i\ne a}
    \frac{\Mcc{ia}}{\Msq}\aP_i(x)-\tP^{q,a}(x)\ell_{ia}\Big) \Big]\,,\\
\varref{usr\_wgts[3]} &= \Msq\Phi
  \frac1{x'}\Big[ \oK^{q,a}(x')+\sum_{i\ne a} 
      \Big(  \frac{\Mcc{ia}}{\Msq}\delta^{qa}\frac{\gamma_i}{1-x'}
             + \frac{\Mcc{ai}}{\Msq} P^{q,a}(x')\ell_{ia} \Big) \Big]\,,\\
\varref{usr\_wgts[4]} &= \Msq\Phi
  \Big[ \tK^{g,a}(x)+\Big( \delta^{ga}\sum_{i\ne a}
    \frac{\Mcc{ia}}{\Msq} \aP_i(x)-\tP^{g,a}(x)\ell_{ia} \Big) \Big]\,,\\
\varref{usr\_wgts[5]} &= \Msq\Phi
  \frac{1}{x'}\Big[ \oK^{g,a}(x')+\sum_{i\ne a}
         \Big( \frac{\Mcc{ia}}{\Msq} \delta^{ga}\frac{\gamma_i}{1-x'}
      + \frac{\Mcc{ai}}{\Msq}P^{g,a}(x')\ell_{ia} \Big) \Big]\,,\\
\varref{usr\_wgts[6]} &= \Msq\Phi
  \Big[ \tK^{q,b}(x)+\Big( \delta^{qb}\sum_{i\ne b}
    \frac{\Mcc{ib}}{\Msq}\aP_i(x)-\tP^{q,b}(x)\ell_{ib} \Big) \Big]\,,\\
\varref{usr\_wgts[7]} &= \Msq\Phi
  \frac1{x'}\Big[ \oK^{q,b}(x')+\sum_{i\ne b} 
             \Big( \frac{\Mcc{ib}}{\Msq}\delta^{qb} \frac{\gamma_i}{1-x'}
         +\frac{\Mcc{bi}}{\Msq}P^{q,b}(x')\ell_{ib} \Big) \Big]\,,\\
\varref{usr\_wgts[8]} &= \Msq\Phi
  \Big[ \tK^{g,b}(x)+\Big( \delta^{gb}\sum_{i\ne b}
    \frac{\Mcc{ib}}{\Msq}\aP_i(x)-\tP^{g,b}(x)\ell_{ib} \Big) \Big]\,,\\
\varref{usr\_wgts[9]} &= \Msq\Phi
  \frac{1}{x'}\Big[ \oK^{g,b}(x')+\sum_{i\ne b}
      \Big(  \frac{\Mcc{ib}}{\Msq}\delta^{gb}\frac{\gamma_i}{1-x'}
          + \frac{\Mcc{bi}}{\Msq} P^{g,b}(x')\ell_{ib} \Big) \Big]\,,\\
\varref{usr\_wgts[10]} &= \Msq\Phi\tP^{q,a}(x)\,,
&&\hskip -64mm
  \varref{usr\_wgts[14]} = \Msq\Phi\tP^{q,b}(x)\,,\\
\varref{usr\_wgts[11]} &= \Msq\Phi\frac{1}{x'}P^{q,a}(x')\,,
&&\hskip -64mm
  \varref{usr\_wgts[15]} = \Msq\Phi\frac{1}{x'}P^{q,b}(x')\,,\\
\varref{usr\_wgts[12]} &= \Msq\Phi\tP^{g,a}(x)\,,
&&\hskip -64mm
  \varref{usr\_wgts[16]} = \Msq\Phi\tP^{g,b}(x)\,,\\
\varref{usr\_wgts[13]} &= \Msq\Phi\frac{1}{x'}P^{g,a}(x')\,,
&&\hskip -64mm
  \varref{usr\_wgts[17]} = \Msq\Phi\frac{1}{x'}P^{g,b}(x')\,.
\end{alignedat}
\end{equation}
In these expressions, $\Phi$ is again the phase-space measure at
the configuration for the event, including Jacobian factors from
any remappings used by the phase-space generator;
$n_B$ is the order in $\alpha_s$ of the Born process.  In
these equations, $a$ is the initial-state parton type and flavor
in hadron 1, and $b$ is the initial-state type and flavor of
the parton from hadron 2.   In the weights where $a$ appears 
(2--5 and 10--13), $x$ is \varref{x1} and $x'$ is \varref{x1p},
whereas in the weights where $b$ appears (6--9 and 14--17),
$x$ is \varref{x2} and $x'$ is \varref{x2p}.

With the additional \varref{usr\_wgts} coefficients, an analysis code
can use the following formulas to recompute the event weight,
\begin{eqnarray}
n &=& \varref{alphas\_power}\,,\\
l &=& \ln\left(\frac{\mur^2}{\varref{ren\_scale}^2}\right) \,, \\
\omega_0 &=& \varref{me\_wgt}+l \, \varref{usr\_wgts[0]}
+ \frac{l^2}{2} \, \varref{usr\_wgts[1]} \,, \\
\omega_i &=& \varref{usr\_wgts[}i\varref{+1]}
+ \varref{usr\_wgts[}i\varref{+9]}
  \ln\left(\frac{\muf^2}{\varref{fac\_scale}^2}\right) \,,\\
m &=& \omega_0\,
   f_1(\varref{id1},\varref{x1},\muf)
\, f_2(\varref{id2},\varref{x2},\muf)
\nonumber\\
&&+\biggl(\sum_{j=1}^4 
   f_1^{(j)}(\varref{id1},\varref{x1},\varref{x1p},\muf)\, \omega_j\biggr)\,
 f_2(\varref{id2},\varref{x2},\muf)\\
&&+f_1(\varref{id1},\varref{x1},\muf)
\biggl(\sum_{j=1}^4 
   f_2^{(j)}(\varref{id2},\varref{x2},\varref{x2p},\muf)\, \omega_{j+4}\biggr)
  \,, \nonumber\\
w &=& m\ \frac{\alpha_s(\mur)^n}{(\varref{alphas})^n} \,,
\end{eqnarray}
where ($r=1$ or~$2$)
\begin{eqnarray}
f_r^{(1)}(i,x,x',\muf) &=& \left\{\begin{array}{l@{\hskip 0pt}cl}
  i = \mbox{quark} &:&f_r(i,x,\muf) \,, \\
  i = \mbox{gluon}&:&\sum_{{\rm quarks}\ q} f_r(q,x,\muf)
   \vphantom{\displaystyle\sum} \,,
                 \end{array}\right.
\\
f_r^{(2)}(i,x,x',\muf) &=& \left\{\begin{array}{l@{\hskip 0pt}cl}
          i = \mbox{quark} &:&
                  f_r(i,x/x',\muf)/x' \,, \\
          i = \mbox{gluon}&:&\sum_{{\rm quarks}\ q}\vphantom{\displaystyle\sum}
                  f_r(q,x/x',\muf)/x' \,,
              \end{array}\right.\\
f_r^{(3)}(i,x,x',\muf) &=& f_r(g,x,\muf) \,, \\
f_r^{(4)}(i,x,x',\muf)&=& f_r(g,x/x',\muf)/x' \,.
\end{eqnarray}
The sums over quarks
are taken over the quark flavors active at the scale $\muf$, typically five.
In these equations, \varref{ren\_scale} is the reference renormalization
scale $\murz$, and \varref{fac\_scale} is the reference factorization
scale $\mufz$.
(In the case of $W$ or $Z\,+\,n_j$-jet processes, $n$ is $n_j+1$
for this contribution.)

\def\nTupleReader{{\tt nTupleReader\/}}
\section{Library for Reading BHS \nTuple{} Files}
\label{LibrarySection}

This section describes a \CC{} library, \nTupleReader{},
 which provides an easy-to-use
interface to the \ntuple{} files described in the above sections. 

\subsection{Dependencies}

The library depends on \textsc{lhapdf} \cite{LHAPDF}
({\tt http://lhapdf.hepforge.org/}) and 
\Root{} \cite{ROOT} ({\tt http://root.cern.ch/drupal/\/}). 

\subsection{Installation}

The library is an autotools package, and is installed using the usual 
paradigm of {\tt configure; make; make install}.
Along with the usual autotools options, the library's {\tt configure} script offers the following options:
\begin{lstlisting}[basicstyle=\ttfamily,keepspaces=true,columns=fullflexible,escapeinside={(*}{*)}]
  --with-lhapdf-path=(*$\langle${\rm path to \textsc{lhapdf} installation}$\rangle$*) 
                    (* {\it Sets the location of the \textsc{lhapdf} installation;\/}*)
                    (* {\it  this option is needed only if the helper program\/}*)
                    (* {\it {\tt lhapdf-config} is not in the executable search path. \/}*)
  --with-root-path=(*$\langle${\rm path to \Root{} installation}$\rangle$*)
                    (* {\it Sets the location of the root installation;\/} *)
                    (* {\it this option is needed only if the helper program \/}*)
                    (* {\it {\tt root-config} is not in the executable search path. \/}*)
  --enable-pythoninterface
                    (* {\it Compiles the python interface (see section \ref{sec:python}).\/}*)
\end{lstlisting}

To install the library, unpack the {\tt .tar.gz} file,
\begin{lstlisting}[basicstyle=\ttfamily,keepspaces=true,columns=fullflexible,escapeinside={(*}{*)}]
tar -xzf ntuplereader-(*$\langle${\rm version\/}$\rangle$*).tar.gz
\end{lstlisting}
where $\langle${\rm version\/}$\rangle$ is a string like `1.0'.
Configure, with options as needed,
\begin{lstlisting}[basicstyle=\ttfamily,keepspaces=true,columns=fullflexible,escapeinside={(*}{*)}]
cd ntuplereader-(*$\langle${\rm version\/}$\rangle$*)
./configure (*$\langle${\rm options}$\rangle$*)
\end{lstlisting}
and then compile and install,
\begin{lstlisting}[basicstyle=\ttfamily,keepspaces=true,columns=fullflexible,escapeinside={(*}{*)}]
make
make install
\end{lstlisting}
To install in a different location than the standard one, use the
{\tt --prefix\/} option to the {\tt configure\/} script.
\subsection{Usage}
Your source code must include the header file 
\nTupleReader{\tt .h} to use the library. 
Your code may access information in the \ntuple{} files via an object of the 
\nTupleReader{} class. We give an example of a \CC{} program
using the library in section \ref{sec:c++}.

The {\tt \nTupleReader-config\/} script returns the compiler and linker
flags needed to compile a user program and link it to the library. The script is located
in the {\tt bin\/} subdirectory of the installation directory.
Assuming this script is in your executable search path, you would
compile and link a program as follows,
\begin{lstlisting}[basicstyle=\ttfamily,keepspaces=true,columns=fullflexible,escapeinside={(*}{*)}]
CFLAGS=`nTupleReader-config --include`
g++ -c $CFLAGS -o NTRexample.o NTRexample.cpp
LDFLAGS=`nTupleReader-config --libs`
g++ $LDFLAGS -o NTRexample NTRexample.o
\end{lstlisting}

\subsection{The {\tt getInfo} program}\label{sec:getInfo}

The {\tt make\/} step creates a program called {\tt getInfo\/} along
with the library. This program takes one argument, the name of a
\ntuple{} file. The program prints useful information about events in
the file:
\begin{itemize}
\item their center-of-mass energy;
\item their initial-state hadrons;
\item the process that generated them;
\item the part of the NLO calculation to which they contribute;
\item the jet algorithms and parameters allowed in an analysis;
\item the minimum jet transverse-momentum cut allowed;
\item additional generation-level cuts, such as the lepton invariant-mass cut for processes with a $Z$ boson.;
\item the electroweak parameters used;
\item the pdf set used in the \ntuple{} generation.
\end{itemize}

\subsection{Member functions of an \nTupleReader{} object}

This section describes the member functions of the \nTupleReader{} class,
 which allow access to the data in \ntuple{} files.
We give an example of their use in section \ref{sec:c++}.

The member functions are,
\begin{description}

\item[]
\verb|void addFile(const std::string &fileName)|\\
Adds a file to the reader.\\
\null\hspace{7mm}\varref{fileName} {\it is the name of the file. \/}

\item[]
\verb|void addFiles(std::vector<std::string> fileNames)|\\
Adds a list of files to the reader\\
\null\hspace{7mm}\varref{fileNames} {\it is a vector of {\tt std::string\/} containing the names of the files to be added, which will be read in the order given. \/}

\item[]
\verb|double computeWeight(double newFactorizationScale,|\\
\verb|                 double newRenormalizationScale)|\\
Returns the weight ({\tt weight\/}) of the current entry recomputed for the new scales, using the current pdf member number in the current pdf set.\\
\null\hspace{7mm}\varref{newFactorizationScale} {\it is the new factorization scale (in GeV) \/}
\\
\null\hspace{7mm}\varref{newRenormalizationScale} {\it is the new renormalization scale (in GeV) \/}

\item[]
\verb|double computeWeight2(double newFactorizationScale,|\\
\verb|                  double newRenormalizationScale)|\\
Returns the secondary weight ({\tt weight2\/}) of the current entry recomputed for the new scales, using the current pdf member number in the current pdf set. One should use this weight for the real part in order to take into account the correlation between the entry and counter entries.\\
\null\hspace{7mm}\varref{newFactorizationScale} {\it is the new factorization scale (in GeV) \/}
\\
\null\hspace{7mm}\varref{newRenormalizationScale} {\it is the new renormalization scale (in GeV) \/}

\item[]
\verb|short getAlphasPower()|\\
Returns the power of the strong coupling constant in the current entry.
\item[]
\verb|long getEndEntryIndex()|\\
Returns the (1-based) index of the entry at which reading will stop.
\item[]
\verb|double getEnergy(int i)|\\
    Returns the energy of the $i^{\rm th}$ particle in the current entry.\\
\null\hspace{7mm}\varref{i} {\it is a 0-based index; an argument equal to or larger than the number of final state particles will throw an {\tt nTR\_OutOfBounds\/} exception. \/}

\item[]
\verb|void getEntry(long index)|\\
Reads the entry corresponding to the index specified; {\tt nextEntry\/}() will start from that position.\\
\null\hspace{7mm}\varref{index} {\it is the index of the entry to be read. \/}

\item[]
\verb|double getFactorizationScale()|\\
Returns the factorization scale used to compute the weights for the current entry.
\item[]
\verb|int getID()|\\
Returns the ID of the current event.
\item[]
\verb|double getId1()|\\
Returns the PDG code for the first (forward) incoming parton in the current entry.
\item[]
\verb|double getId2()|\\
Returns the PDG code for the second (backward) incoming parton in the current entry.
\item[]
\verb|long getIndexOfNextEntry()|\\
Returns the index of the next entry.
\item[]
\verb|double getMEWeight()|\\
Returns the weight for the current entry omitting pdf factors.
\item[]
\verb|double getMEWeight2()|\\
Returns the secondary weight for the current entry omitting the pdf factors, to be used as described in \sect{NTupleUseSection} to obtain the correct estimate of the statistical uncertainty.
\item[]
\verb|long getMaxEvent()|\\
Returns the ID of the event (phase-space configuration--counter-configuration group of entries, or simply entries for contribution types with one entry per event) after which reading will stop.
\item[]
\verb|long getNumberOfEntries()|\\
Returns the total number of entries.
\item[]
\verb|int getPDGcode(int i)|\\
    Returns the PDG code of the $i^{\rm th}$ particle in the current entry.\\
\null\hspace{7mm}\varref{i} {\it is a 0-based index; an argument equal to or larger than the number of final state particles will throw an {\tt nTR\_OutOfBound\/} exception. \/}

\item[]
\verb|int getParticleNumber()|\\
Returns the number of final state particles in the current entry.
\item[]
\verb|double getRenormalizationScale()|\\
Returns the renormalization scale used to compute the weights for the current entry.
\item[]
\verb|long getStartEntryIndex()|\\
Returns the (1-based) index of the entry at which reading will start.
\item[]
\verb|char getType()|\\
    Returns the type of the current entry, `B' standing for born, `I' for integrated subtraction, `V' for the virtual, and `R' for the subtracted real emission.
\item[]
\verb|double getWeight()|\\
Returns the weight ({\tt weight\/}) for the current entry.
\item[]
\verb|double getWeight2()|\\
Returns the secondary weight ({\tt weight2\/}) for the current entry, to be used as described in \sect{NTupleUseSection} to obtain the correct estimate of the statistical uncertainty.
\item[]
\verb|double getX(int i)|\\
    Returns the $x$ component of the $i^{\rm th}$ particle's momentum in the current entry.\\
\null\hspace{7mm}\varref{i} {\it is a 0-based index; an argument equal to or larger than the number of final state particles will throw an {\tt nTR\_OutOfBound\/} exception. \/}

\item[]
\verb|double getX1()|\\
Returns the momentum fraction $x_1$ in the current entry.
\item[]
\verb|double getX2()|\\
Returns the momentum fraction $x_2$ in the current entry.
\item[]
\verb|double getY(int i)|\\
    Returns the $y$ component of the $i^{\rm th}$ particle's momentum in the current entry.\\
\null\hspace{7mm}\varref{i} {\it is a 0-based index; an argument equal to or larger than the number of final state particles will throw an {\tt nTR\_OutOfBound\/} exception. \/}

\item[]
\verb|double getZ(int i)|\\
    Returns the $z$ component of the $i^{\rm th}$ particle's momentum in the current entry.\\
\null\hspace{7mm}\varref{i} {\it is a 0-based index; an argument equal to or larger than the number of final state particles will throw an {\tt nTR\_OutOfBound\/} exception. \/}

\item[]
\verb|bool nextEntry()|\\
Reads the next entry and returns true upon success, false otherwise (including when the end of the file is reached).
\item[]
\verb|void setEndEntryIndex(long index)|\\
Sets the index of the entry at which reading will stop.\\
\null\hspace{7mm}\varref{index} {\it 1-based index at which the reading will stop (the \varref{index}-th entry will not be read) \/}

\item[]
\verb|void setMaxEvent(long count)|\\
Sets the reader to stop reading entries when the given number of events (phase-space configuration--counter-configuration groups of entries, or simply entries for contribution types with one entry per event) have been read.\\
\null\hspace{7mm}\varref{count} {\it is a 1-based sequence number specifying the first event that will not be read. \/}

\item[]
\verb|void setPDF(const std::string &name)|\\
Sets the pdf set to be used.\\
\null\hspace{7mm}\varref{name} {\it is the name of the file to be loaded by \textsc{LHAPDF}, for example {\tt CT10.LHgrid\/}. \/}

\item[]
\verb|void setPDFmember(int member)|\\
Sets the pdf member number to be used.\\
\null\hspace{7mm}\varref{member} {\it is an integer labeling the member; 0 is typically used to denote the central value. \/}

\item[]
\verb|void setPP()|\\
Sets the initial state to proton--proton. This is the default if no calls to {\tt setPP()} or {\tt setPPbar()} are issued. This routine should only be invoked before using files generated for proton--proton colliders.
\item[]
\verb|void setPPbar()|\\
Sets the initial state to proton--antiproton. This routine should only be invoked before using files generated for proton--antiproton colliders.
\item[]
\verb|void setStartEntryIndex(long index)|\\
Sets the index of the entry at which reading will start.\\
\null\hspace{7mm}\varref{index} {\it 1-based index at which the reading will start (the \varref{index}-th entry will be read at the next call of {\tt nextEntry()}) \/}

\item[]
\verb|void setStartEvent(long count)|\\
Sets the number for the event sequence counter at which the library will start reading in events (phase-space configuration--counter-configuration groups of entries, or simply entries for contribution types with one entry per event).\\
\null\hspace{7mm}\varref{count} {\it is a 1-based sequence number specifying the first event that will be read. \/}

\end{description}
\subsection{Histogram Implementation Example}
\label{sec:histogram}
This section gives an example of the implementation of a histogramming
procedure, as described in section~\ref{NTupleUseSection}.

\noindent\rule{15cm}{1pt}  
\begin{lstlisting}[basicstyle=\ttfamily,keepspaces=true,columns=fullflexible,escapeinside={(*}{*)}]
nTupleReader r;
(*$\vdots$*)
bool notFinished=r.nextEntry();
int lastID=r.getID();

while(notFinished){  
  double wgt;
  // compute the value of the weight wgt here...
  int ID=r.getID()  
  int bin=findBin(x);
  finalHistogram[bin]+=wgt;  
  tempHistogram[bin]+=wgt;
    
  if (ID != lastID){ 
    for (int bin=0;bin<NbrBins;bin++){
      weightsSquare[bin]+=tempHistogram[bin]*tempHistogram[bin];
      if (   tempHistogram[bin] != 0.0 ){
        NbrEntries[bin]++;
      }
      tempHistogram[bin]=0.;
    }
    lastID=id;
  }
  notFinished=r.nextEvent()
}
\end{lstlisting}
\rule{15cm}{1pt} \\ 
where \varref{NbrBins} is the numbers of bins in the histograms.
 
%%%%%%%%%%%%%%%%%%%
\subsection{Example of the usage of the python interface}\label{sec:python}
To use the python interface, make sure that the installation path for the
libraries installed with the \CC{} library is included in the locations
searched for python modules.  The following is a sample python program,
\\\rule{15cm}{1pt}  
\begin{lstlisting}[basicstyle=\ttfamily,keepspaces=true,columns=fullflexible,escapeinside={(*}{*)}]
import nTupleReader as NR

r=NR.nTupleReader()
r.addFile('@prefix@/share/ntuplereader/sample.root')

r.nextEntry()

for i in range(r.getParticleNumber()):
    print "p(%d)=(%f,%f,%f,%f)" % (
              i,
              r.getEnergy(i),
              r.getX(i),
              r.getY(i),
              r.getZ(i)
         )
\end{lstlisting}
\rule{15cm}{1pt}  
\subsection{Example of the usage of the library in a 
C++ program}\label{sec:c++}

The following listing shows an example of a \CC{} program using the 
\nTupleReader{} library. This example is included after installation
in the directory {\tt share/ntuplereader/\/}.
\\\rule{15cm}{1pt}  
\begin{lstlisting}[basicstyle=\ttfamily,keepspaces=true,columns=fullflexible,escapeinside={(*}{*)}]
#include "nTupleReader.h"

using namespace std;

int main(){

	std::vector<std::string> fs;
	fs.push_back("@prefix@/share/ntuplereader/sample.root");

	nTupleReader r;


	r.setPDF("cteq6ll.LHpdf");
	r.addFiles(fs);

	while(r.nextEntry()){
		int id=r.getID();
		cout << 
                   "Checking momentum conservation for event ID: " 
				<< id << std::endl;
		double sumX=0;
		double sumY=0;
		double sumZ=0;
		int nbrP=r.getParticleNumber();
		std::cout << "Number of particles: " 
                          << nbrP << std::endl;
		for (int i=0;i<nbrP;i++){
			sumX+=r.getX(i);
			sumY+=r.getY(i);
			sumZ+=r.getZ(i);
		}
		std::cout << "Sum X: " << sumX << std::endl;
		std::cout << "Sum Y: " << sumY << std::endl;
		std::cout << "Sum Z: " << sumZ 
                          << " to be compared with: " 
                          <<  3500 * (r.getX1()-r.getX2())
                          <<  std::endl; 
	}

	return 0;

}
\end{lstlisting}
\rule{15cm}{1pt}  

\section{Conclusions}
\label{Conclusions}

In this Article we described software tools for obtaining 
predictions to NLO in QCD
using an \ntuple{} event-file format generated by
\Sherpa{} using the \BlackHat{} software library.  
This framework offers a convenient means for evaluating
cross-sections and distributions with different parton distribution
functions, bypassing repeated and computationally costly evaluations
of matrix elements. The set-up described here makes it straightforward
to compute scale-variation bands and to obtain uncertainty estimates
due to imprecise knowledge of the parton distribution functions. The
\ntuple{} files also make it convenient to study the effects of
varying jet algorithms and cuts, within a wide range of
commonly-used ones.  These tools have already proved useful in a
number of theoretical~\cite{pureQCD, photonZ3,W5j} and experimental
studies~\cite{ATLASZJets,ATLASWJets}. 
The reader may find an up-to-date list of available \ntuple{} sets
at {\tt http://blackhat.hepforge.org/trac/wiki/Availability\/}, stored
at locations given in {\tt http://blackhat.hepforge.org/trac/wiki/Location}.
We look forward to further
theoretical and experimental studies using the tools described here.

\section*{Acknowledgments}

We thank Tanju Gleisberg for his collaboration in the initial stages
of the development described above; he contributed
in an important way to the tools described here.
We are also grateful to Kemal
Ozeren for his collaboration and input, and
we thank Joey Huston for his continual encouragement.  We also thank
David Saltzberg, and 
Eric Takasugi for discussions and
feedback.
This research was supported by the US Department of Energy under
contracts DE--AC02--76SF00515 and DE--FG02--13ER42022.  DAK’s research
is supported by the European Research Council under Advanced
Investigator Grant ERC--AdG--228301.  
DM's work was supported by the
Research Executive Agency (REA) of the European Union under the Grant
Agreement number PITN--GA--2010--264564 (LHCPhenoNet). SH's
work was partly supported by a grant from the US LHC Theory
Initiative through NSF contract PHY--0705682.  This research used
resources of Academic Technology Services at UCLA, and of the National
Energy Research Scientific Computing Center, which is supported by the
Office of Science of the U.S. Department of Energy under Contract
No.~DE--AC02--05CH11231.

%%%%%%%%%%%%%%%%%%%%%%%%%%%%%%%%

\appendix

\section{Tables of Cross Section Values}

In this appendix we provide tables of cross sections obtained using the
$\sqrt{s}=8$~TeV \ntuples{} currently available on CASTOR and on the LHC Grid.  
These can serve
as a reference for comparison purposes, and as a cross check
that the sum over different contributions has been carried out
correctly in user code.
  We list cross sections for a
specific set of cuts identical to the ones used at 7 TeV in
refs.~\cite{W4,Z4,pureQCD}.
(Sample distributions listed in \Tab{tab:histograms}
  are included with each \ntuple{} file.)

\subsection{Inclusive \Wjn-jet Production}

%%%%%%%%%%%%%%%%%%%%%%%% TABLE %%%%%%%%%%%%%%%%%%%%%%
\begin{table}[thb]
\centering
\begin{tabular}{||c||c|c||c|c||}
\hline
Process &  $W^-$ LO  & $W^-$ NLO  & $W^+$ LO & $W^+$ NLO  \\ 
\hline
\Wj  & $341.4(0.2)$ & $422.3(0.6)$ & $487.4(0.4)$ & $597(2)$\\ \hline
\Wjj  & $105.1(0.1)$ & $ 104.1(0.3)$ & $158.1(0.2)$ & $154.0(0.5)$\\ \hline
\Wjjj  & $27.6(0.04)$ & $23.9 (0.1)$ & $43.85(0.08)$ & $37.2(0.4)$\\ \hline
\end{tabular}
\caption{Total cross sections in pb for \Wjn{} jet production at the
LHC at $\sqrt{s}=8$~TeV, using the  anti-$k_T$ jet algorithm with  $R=0.5$.
}
\label{CrossSectionW-Anti-kt-R5Table}
\end{table}
%%%%%%%%%%%%%%%%%%%%%%%%%%%%%%%%

We display the $\sqrt{s} = 8$ TeV cross sections for $W+n$ jet production in
\Tab{CrossSectionW-Anti-kt-R5Table}.  We
define jets using the anti-$k_T$ algorithm~\cite{AntiKT} with
parameter $R = 0.5$.  We apply the following cuts,
\begin{eqnarray}
&& \ET^{e} > 20 \hbox{ GeV} \,, \hskip 1.5 cm 
|\eta^e| < 2.5\,, \hskip 1.5 cm 
\ETsl > 20 \hbox{ GeV}\,,  \hskip 1.5 cm \nn\\
&& \pT^\jet > 25 \hbox{ GeV}\,, \hskip 1.5 cm 
|\eta^\jet|<3\,,  \hskip 1.5 cm M_{\rm T}^W > 20  \hbox{ GeV}\,. 
\end{eqnarray}
The transverse mass of the $W$-boson is computed from the
kinematics of its decay products, $W\rightarrow e\nu_e$:
 $M_{\rm T}^W=\sqrt{2 \ET^e
   \ET^\nu (1- \cos(\Delta \phi_{e\nu}))}$.
In this case, the factorization and renormalization scales are
set to,
\begin{equation}
\murz = \mufz = {\textstyle\frac12}\HTpartonicp\,,
\end{equation}
where $\HTpartonicp$ is defined in \eqn{HTpartonicp}.  
The LO cross sections are computed using 
MSTW2008~\cite{MSTW2008} LO pdfs, and the NLO cross sections
using the MSTW2008 NLO pdfs.  In each case, we use
the $\alpha_s(\mu)$ corresponding to the parton distribution set.
  The corresponding 7 TeV cross sections
are given in table~I of ref.~\cite{W5j}.

\subsection{Inclusive \Zgamjn-jet Production}

%%%%%%%%%%%%%%%%%%%%%%%% TABLE %%%%%%%%%%%%%%%%%%%%%%
\begin{table}[thb]
\centering
\begin{tabular}{||c||c|c||}
\hline
Process & LO  &  NLO     
\\ \hline
\Zj &$84.05(0.04)^{}_{}$ & $102.6(0.2)^{}_{}$ 
\\ \hline
\Zjj &$26.86(0.02)^{}_{}$ & $26.50(0.06)^{}_{}$ 
\\ \hline
\Zjjj & $7.452(0.009)^{}_{}$ & $6.58(0.09)^{}_{}$ 
\\ \hline
\end{tabular}
\caption{Total cross sections in pb for \Zgamjn{} jet production at the
LHC at $\sqrt{s}=8$~TeV, using the  anti-$k_T$ jet algorithm with $R=0.5$
and the cuts given in \eqn{Zcuts}. 
}
\label{CrossSectionZ-Anti-kt-R5Table}
\end{table}
%%%%%%%%%%%%%%%%%%%%%%%%%%%%%%%%

\def\ET{E_{\rm T}}
For the 8 TeV cross sections listed in
\Tab{CrossSectionZ-Anti-kt-R5Table}, we choose the anti-$k_T$
algorithm with $R=0.5$ and impose the following set of cuts,
\begin{eqnarray}
&& \ET^{e} > 20 \hbox{ GeV} \,, \hskip 1.5 cm 
|\eta^e| < 2.5\,, \hskip 1.5 cm 
66\hbox{ GeV}<M_{e^+ e^-}<116\hbox{ GeV}\,, \nn\\
&& \pT^\jet > 25 \hbox{ GeV}\,, \hskip 1.5 cm |\eta^\jet|<3\,,
\label{Zcuts}
\end{eqnarray}
where $\eta$ is the pseudorapidity and $M_{e^+ e^-}$ is the invariant
mass of the $e^+ e^-$ decay pair.  The invariant-mass cut is
tighter than that used in generating the \ntuples{}, and other
cuts are added in this calculation, illustrating precisely
this flexibility of
the \ntuple{} setup.
The factorization and renormalization
scales are set to
\begin{equation}
\murz = \mufz = \mbox{\footnotesize $\displaystyle \frac12$}\HTpartonicp\,,
\end{equation}
where $\HTpartonicp$ is defined in \eqn{HTpartonicp}. 
The MSTW2008 parton distribution functions are used.
Cross sections at 7 TeV are given in ref.~\cite{Z4}.

\subsection{Inclusive $n$-jet Production}

%%%%%%%%%%%%%%%%%%%%%%%% TABLE %%%%%%%%%%%%%%%%%%%%%%
\begin{table}[hbt]
\centering
\begin{tabular}{||c||c|c||}
\hline
Jets & LO  &  NLO     
\\ \hline
2 &$1232.5(0.2)$ & $1526.(2)$
\\ \hline
3 &$126.74(.03)$ & $71.9(0.3)$
\\ \hline
4 & $14.36(0.01)$ & $8.12(0.17)$ 
\\ \hline
\end{tabular}
\caption{Total cross sections in nb for pure jet production at the LHC
  at $\sqrt{s}=8$~TeV, using the anti-$k_T$ jet algorithm with $R=0.4$
  and the cuts given in \eqn{JetCuts}. }
\label{CrossSectionJets-Anti-kt-R4Table}
\end{table}
%%%%%%%%%%%%%%%%%%%%%%%%%%%%%%%%

The cross sections for $2,3,4$-jet production at $\sqrt{s} = 8$ TeV
are given in \Tab{CrossSectionJets-Anti-kt-R4Table}.  In this
case we use the anti-$k_T$ jet algorithm with $R=0.4$ with the same
cuts as in ref.~\cite{pureQCD}:
\begin{eqnarray}
\pT^\jet > 60   \hbox{ GeV} \,, \hskip 1.5 cm 
\pT^{\rm leading\ jet} > 80   \hbox{ GeV} \,, \hskip 1.5 cm 
|y|^\jet < 2.8\,,
\label{JetCuts}
\end{eqnarray}
where $\pT^{\rm leading\ jet}$ is the transverse momentum of the
leading jet ordered in transverse momentum and $y^\jet$ is the
rapidity of a jet.  In this case, the factorization and
renormalization scale are chosen to be,
\begin{equation}
\murz = \mufz \equiv \mu =  
\mbox{\footnotesize $\displaystyle \frac12$}\sum_j p_T^j\,,
\end{equation}
where the sum runs over all final-state partons $j$. Again we use 
the MSTW2008 parton distribution functions.
 The 7 TeV cross
sections are given in table~I of ref.~\cite{pureQCD}.
\FloatBarrier

%%%%%%%%%%%%%%%%%%%%%%%%%%%%%%%%%%%%%%%%%%%%%%%%%
\bibliographystyle{elsarticle-num}
\bibliography{ntuples}






% \end{thebibliography}

\end{document}